\documentclass[aps,prd,nofootinbib,twocolumn,superscriptaddress,preprintnumbers,balancelastpage,longbibliography,nobibnotes]{revtex4-1}

\usepackage{xcolor}
\usepackage{rotating} 
\usepackage{amsmath,amssymb,mathtools,bm}
\usepackage{graphicx, color, hepunits}
\usepackage[colorlinks=true
,urlcolor=blue
,anchorcolor=blue
,citecolor=blue
,filecolor=magenta
,linkcolor=red
,menucolor=blue
,linktocpage=true
,pdfproducer=medialab
,pdfa=true
]{hyperref}
\usepackage[utf8]{inputenc}
\usepackage[english]{babel}

\newcommand{\mA}{m_{A'}}

\newcommand{\pp}{\textbf{p}}
\newcommand{\ppr}{\textbf{p}'}
\newcommand{\kk}{\textbf{k}}
\newcommand{\pf}{\textbf{p$_f$}}
\newcommand{\q}{\textbf{q}}
\newcommand{\V}{\textbf{V}}

\begin{document}

\title{Dark Photons from Charged Pion Bremsstrahlung at Proton Beam Experiments}

\author{David Curtin}
\affiliation{Department of Physics, University of Toronto, Toronto, ON M5S 1A7, Canada}

\author{Yonatan Kahn}
\affiliation{Department of Physics, University of Illinois Urbana-Champaign, Urbana, IL 61801, U.S.A.}
\affiliation{Illinois Center for Advanced Studies of the Universe, University of Illinois Urbana-Champaign, Urbana, IL 61801, U.S.A.}

\author{Rachel Nguyen}
\affiliation{Department of Physics, University of Illinois Urbana-Champaign, Urbana, IL 61801, U.S.A.}
\affiliation{Illinois Center for Advanced Studies of the Universe, University of Illinois Urbana-Champaign, Urbana, IL 61801, U.S.A.}

\begin{abstract}
The production and subsequent re-scattering of secondary pions produced in proton beam dumps provides additional opportunities for the production of light new particles like dark photons. 
This new mechanism has been overlooked in the past but can 
extend the mass reach of the SpinQuest experiment and its proposed DarkQuest upgrade.
We use chiral perturbation theory
to calculate the production of kinetically mixed dark photons through bremsstrahlung off secondary charged pions.
We find that the reach of SpinQuest/DarkQuest can be pushed further into the multi-GeV mass range compared to estimates based only on primary dark photon production through meson decay or proton bremsstrahlung.
Our analysis can be regarded as the first of several steps to include secondary pion contributions. In an upcoming analysis we will extend our calculation into the high-momentum-transfer regime through the use of pion PDFs and including hadronic resonances, which will further increase the estimated mass reach.
\end{abstract}
\maketitle

\section{Introduction}

A simple, renormalizable extension to the Standard Model (SM) is a new massive U(1) gauge boson $A'$, dubbed a dark photon~\cite{Fayet:1980rr,Okun:1982xi,Georgi:1983sy,Holdom:1985ag,Fayet:1990wx}. The dark photon can kinetically mix with the SM photon with strength $\epsilon \ll 1$, after which particles with charge $Q$ under electromagnetism, even if neutral under the new U(1), can acquire small couplings $\epsilon Q$ to the dark photon. This observation has opened a large number of experimental avenues to pursue the dark photon (which may mediate dark matter interactions, or which may itself be the dark matter) in the laboratory; see Ref.~\cite{Fabbrichesi:2020wbt} for a review.

While dark photons may be produced at colliders, ``intensity frontier'' experiments such as beam dumps have emerged as a promising search strategy for relatively low-mass dark photons, largely due to their enormous luminosity~\cite{Batell:2022dpx,Gori:2022vri}. To date, the dominant production mechanisms which have been considered are neutral meson decay~\cite{Batell:2009di}, for example $\pi^0 \to \gamma A'$ where the dark photon replaces an ordinary photon from the SM process $\pi^0 \to \gamma \gamma$, or  bremsstrahlung~\cite{Bjorken:2009mm}, where the primary charged-particle beam radiates an $A'$ instead of a SM photon. So far, electron and proton beam dump experiments have only been able to constrain the dark photon parameter space up to masses of a few hundred MeV. Many experiments have proposed future searches to search for dark photons up to larger masses $m_{A'} \gtrsim 1 \ {\rm GeV}$ through the mechanisms previously listed; see Refs.~\cite{Batell:2022dpx,Gori:2022vri} and references therein for a comprehensive comparison of different approaches.

In this paper, we point out a previously overlooked mechanism for dark photon production at beam dump experiments that takes advantage of the copious flux of secondary charged particles from hadron beams. Since any charged particle is in principle a source for kinetically-mixed dark photons, we propose searching for dark photons through \emph{secondary charged pion bremsstrahlung} during pion-nuclear scattering in the beam dump.\footnote{We note that this process was recently considered in the astrophysical context of supernovae~\cite{Shin:2022ulh}.} Pions are the most common secondary particle produced by hadronic interactions, but they are lighter than the proton, which enhances the reach to heavy $A'$s. Specifically, the cross section for massive dark photon production peaks when the $A'$ takes all of the available energy from the initial state, so long as the $A'$ is heavier than the emitting particle~\cite{Bjorken:2009mm}. This phenomenon was first noticed in the context of lepton beam experiments, but we find that the same phenomenology holds when we use the chiral Lagrangian to model pion-nucleus interactions. We show that the large flux of pions implies that secondary dark photon production through this mechanism is the dominant source of dark photons for $m_{A'} \gtrsim 1.5 \ {\rm GeV}$. The large $A'$ energies give a long lab-frame decay length and an enhanced probability of decay in the instrumented region of the detector, beyond the beam dump. 

This additional source of $A'$s can extend the search for dark photons at current and proposed beam dump experiments to larger dark photon masses and smaller values of $\epsilon$. 
Primary dark photon searches can be performed at long-lived particle experiments at the Large Hadron Collider~\cite{Beacham:2019nyx}, but in those cases it is not obvious that dark photon production through secondary pion interactions significantly increases reach.
MATHUSLA~\cite{Chou:2016lxi} suffers from the fact that significantly produced GeV-scale dark photons have decay lengths much smaller than the distance from the collision point to the detector, and this issue persists for secondary production from pion bremsstrahlung in the main detector material; 
CODEX-b~\cite{Aielli:2019ivi} could produce dark photons through secondary pion interactions in the shield separating it from the LHCb interaction point, but its smaller size and luminosity make minimal dark photon searches challenging\footnote{Including quark-level production from high-momentum-transfer collisions may change the picture for MATHUSLA and CODEX-b, but we leave this for future investigation.}; and while
FASER~\cite{Feng:2017uoz} is better suited to search for low-mass minimal dark photons, forward charged pions produced in LHC collisions are deflected by the magnets and do not point at FASER once they undergo scattering in material.
By contrast, the SpinQuest 120 GeV proton fixed-target experiment at Fermilab~\cite{Berlin:2018pwi} (and its proposed DarkQuest upgrade~\cite{Apyan:2022tsd}) has the near-optimal geometry to take advantage of secondary pion-induced processes, and thus we focus our analysis on this setup.

This paper is organized as follows. In Section \ref{sec:production}, we review the kinetically-mixed dark photon model, calculate the $A'$ production cross section from pion-nuclear scattering, and estimate the expected number of visibly-decaying dark photons as a function of the experimental geometry. In Section \ref{sec:signals}, we illustrate how this production mechanism extends the sensitivity of SpinQuest and DarkQuest to dark photons of $m_{A'} \sim 2 $ GeV and $\epsilon \sim 10^{-7}$. We conclude in Sec.~\ref{sec:conclusions} with a discussion of the prospects for the currently-operating SpinQuest experiment and future DarkQuest upgrades.

\section{Production and Decay of Dark Photons}
\label{sec:production}

\subsection{Dark photon model}

The kinetically-mixed dark photon model (see Ref.~\cite{Fabbrichesi:2020wbt} for a review) is defined by two parameters, the dimensionless kinetic mixing $\epsilon$ and the dark photon mass $m_{A'}$, through the Lagrangian 
\begin{equation}
    \mathcal{L} \supset -\frac{1}{4} F_{\mu \nu} F^{\mu \nu} - \frac{1}{4} F'_{\mu \nu} F'^{\mu \nu} +\frac{\epsilon}{2} F_{\mu \nu} F'^{\mu \nu} + \frac{1}{2} m_{A'}^2 A'_{\mu} A'^{\mu}, 
\end{equation}
where $F_{\mu \nu} = \partial_{\mu} A_{\nu} - \partial_{\nu} A_{\mu}$ is the electromagnetic field strength and $F'_{\mu \nu} = \partial_{\mu} A'_{\nu} - \partial_{\nu} A'_{\mu}$ is the dark photon field strength. The value of $\epsilon$ is arbitrary but numerous collider searches and precision measurements of the electron magnetic moment constrain $\epsilon \lesssim 10^{-3}$ for $m_{A'}$ in the MeV -- GeV range~\cite{Fabbrichesi:2020wbt}. We can perform a field redefinition of the SM photon,
\begin{equation}
\label{eq:massbasis}
    A_{\mu} \to A_{\mu} -  \epsilon A'_{\mu},
\end{equation}
which diagonalizes the kinetic terms to $\mathcal{O}(\epsilon)$ by eliminating the mixing term. In this basis, where $A'$ is a propagating mass eigenstate, the Lagrangian picks up an additional current $\mathcal{L} \supset -\epsilon J_{\mu} A'^{\mu}$, which couples electromagnetic currents $J_\mu$ to the dark photon. Therefore, any particle with charge $Q$ interacts with a dark photon with strength $\epsilon Q$. In this paper, we focus on charged pions $\pi^\pm$ and protons.

\subsection{Dark photon production cross section}
\label{sec:cs}
\begin{figure}
    \centering
    \includegraphics[width = 0.49\textwidth]{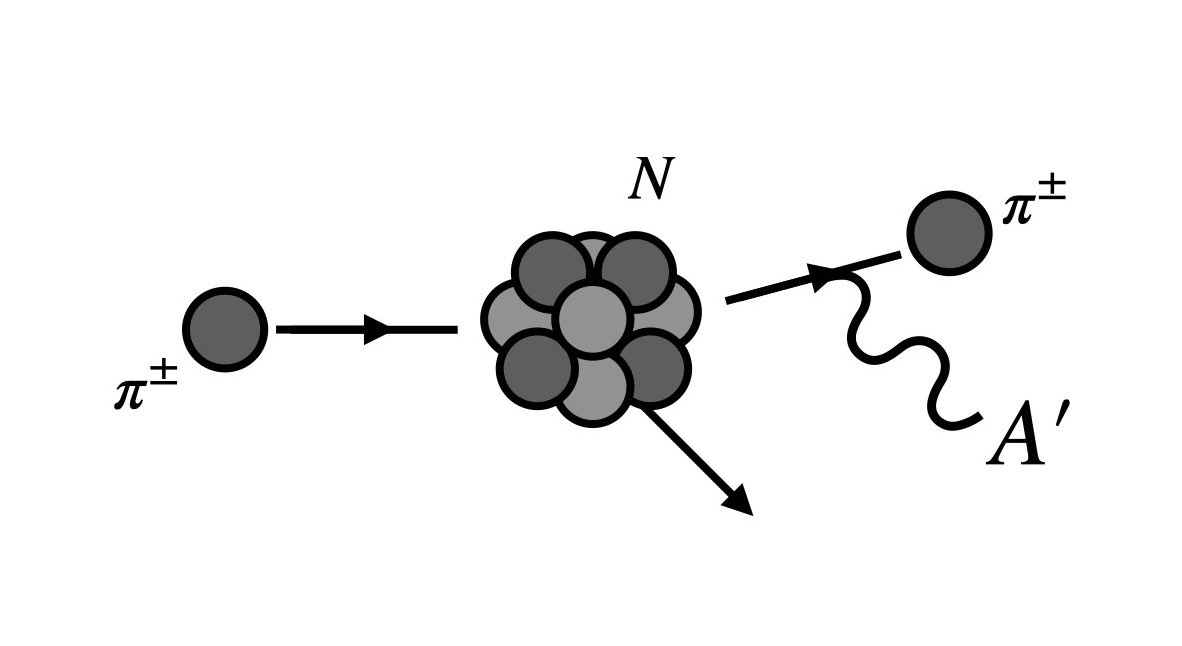}
    \caption{Schematic illustration of a charged pion scattering off a nucleus $N$ and radiating a dark photon $A'$.}
    \label{fig:bremscattering}
\end{figure}
Charged pions may emit dark photons through bremsstrahlung during nuclear scattering, $\pi^\pm N \to \pi^\pm N A'$, as shown in Fig. \ref{fig:bremscattering}. As we discuss in Sec.~\ref{sec:signals} below, there are $\mathcal{O}(10)$ pions produced per primary proton, which effectively increases the luminosity of a proton fixed-target experiment through this scattering channel. The pion-nucleon scattering could occur through either photon exchange or the strong interaction. The cross section for both of these processes will be suppressed by $\epsilon^2$ compared to ordinary bremsstrahlung of a SM photon. However, the QED process has an additional suppression compared to the strong interaction because the $t$-channel exchange of a massless photon yields a cross section which peaks at zero momentum transfer, while production of a massive dark photon is maximized with a momentum transfer on the order of $m_A'$. By contrast, the short-range strong interactions have typical momentum transfer on the order of the pion decay constant, $f_\pi \simeq 93 \ {\rm MeV}$, which is well matched to heavy dark photons. In Appendix~\ref{app:QEDsuppress} we provide more detail on the QED cross section, but for what follows we will focus exclusively on the strong interaction process.

At momentum transfers well below $4\pi f_\pi \simeq 1.2 \ {\rm GeV}$, pion-nuclear scattering is well-described by chiral perturbation theory. However, dark photons heavier than 1.2 GeV can be produced with momentum transfers exceeding the effective field theory cutoff, and thus a full calculation of the production rate would require a quark-level description. As a conservative estimate for the $A'$ production cross section, we will work exclusively within chiral perturbation theory, cutting off all momentum integrals at $4\pi f_\pi$. This artificially suppresses the dark photon production rate at large masses. However, we expect the sensitivity to dark photons to continue to improve at larger $m_{A'}$ from computing the scattering rate using the pion parton distribution functions (PDFs) at the appropriate energy scale~\cite{Bourrely:2018yck}. We leave this study to future work.

\subsubsection{Chiral Perturbation Theory}

\begin{figure*}[t]
    \centering
    \includegraphics[width = 0.48\textwidth]{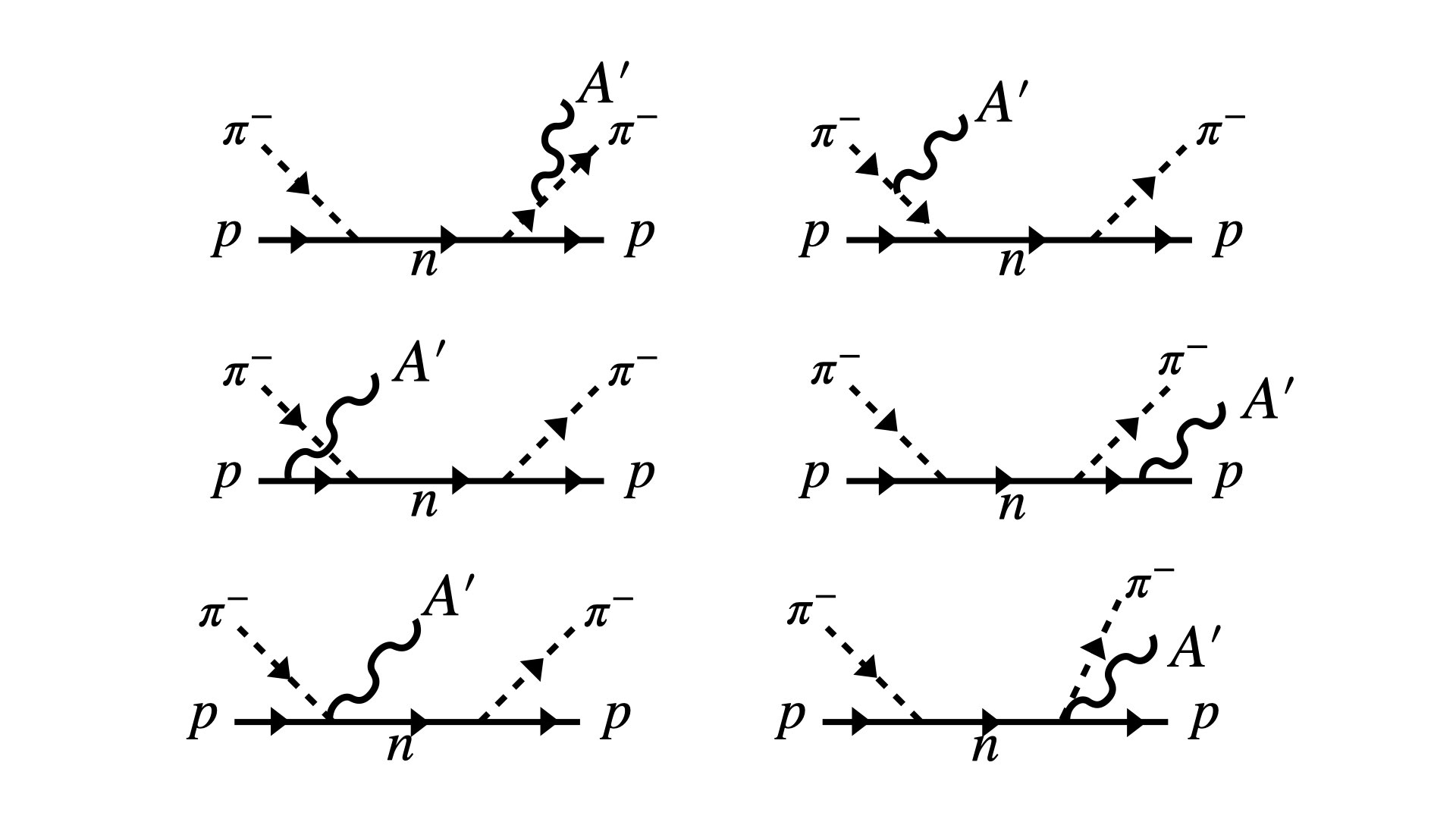}
    \hspace{0.2cm}
    \includegraphics[width = 0.48\textwidth]{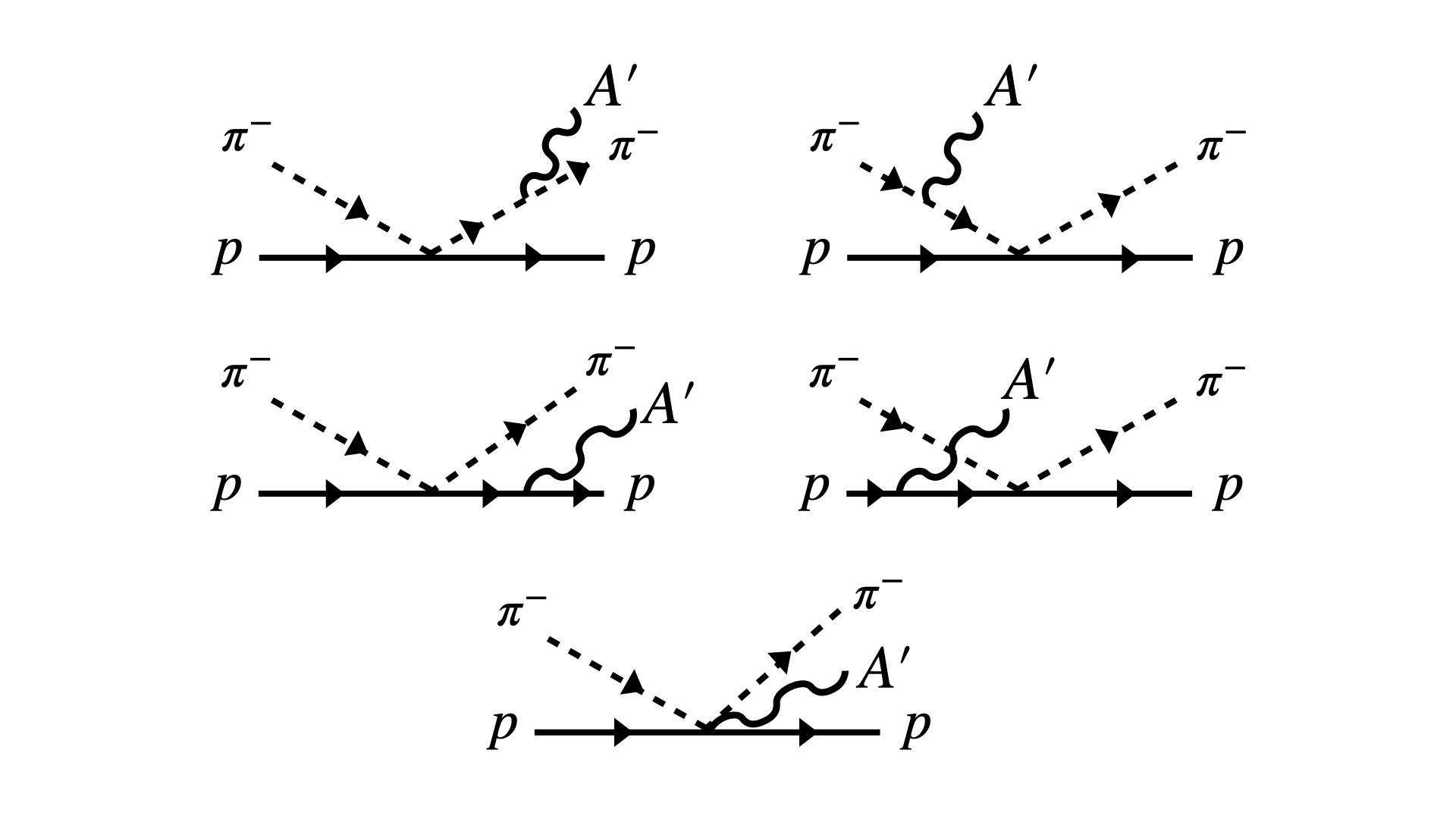}
    \caption{Leading-order Feynman diagrams for $A'$ bremsstrahlung in charged pion-proton scattering. These $s$-channel diagrams correspond specifically to $\pi^-$-proton scattering, but a similar set of $t$-channel diagrams exists for $\pi^+$-proton scattering. Likewise, a similar set of diagrams describes pion-neutron scattering, but in that case the $A'$ can only be emitted from an internal proton line or the external pion lines. The pion-nucleus cross section is obtained by summing incoherently over the proton and neutron contributions, Eq.~(\ref{eq:totalcs}). The left set of diagrams corresponds to scattering from the axial vector term (\ref{eq:axial}) in the chiral Lagrangian, where each pion-nucleon vertex has an odd number of pion lines, while the right set of diagrams correspond to the polar vector term (\ref{eq:vector}) with an even number of pions. The 4- and 5-point contact interactions involving the $A'$ are necessary to satisfy the Ward identity.}
    \label{fig:ChiralFeynmanDiags}
\end{figure*}

To leading order in derivatives, the two-flavor isospin-symmetric chiral Lagrangian is
\begin{align}
    \mathcal{L_{\chi}} = \frac{f_{\pi}^2}{4} \text{Tr}( D_{\mu} U^{\dagger} D^{\mu} U) + \frac{1}{4} m_{\pi}^2 f_{\pi}^2 \text{Tr}(U + U^{\dagger} - 2),
\end{align}
where
\begin{align}
    U(x) = \exp \left[ \frac{i}{f_{\pi}}\pi^a(x)  \sigma^a \right]
\end{align}
is an SU(2) matrix containing the pion fields $\pi^0 = \pi^3$ and $\pi^{\pm} = \frac{1}{\sqrt{2}} (\pi^1 \mp i \pi^2)$, $\sigma^a$ ($a = 1, 2, 3$) are the Pauli matrices.
We define the QED covariant derivative by 
\begin{equation}
    D_{\mu} U = \partial_\mu U - i v_{\mu} U + i U v_{\mu},
\end{equation}
where $v_{\mu} = - e A_{\mu} \sigma^3 / 2$ provides the QED coupling.

To add pion-nucleon interactions, we follow the formalism of Refs.~\cite{Scherer:2002tk,Shin:2022ulh,Ellis:1997kc,Bernard:1995gx,Bernard:1997tq}, which introduces an axial vector field, $u_{\mu}$, and a connection, $\Gamma_{\mu}$, defined as
\begin{align}
    u_{\mu} =&-i u^{\dag} (D_{\mu}U) u^{\dag},\\
    \Gamma_{\mu} &\equiv \frac{1}{2}\left[u ( \partial_{\mu} - i v_{\mu}) u^{\dag} + u^{\dag} ( \partial_{\mu} - i v_{\mu}) u \right ],
\end{align}
where $u^2 = U(x)$. The connection couples to the nucleons through the chiral covariant derivative
\begin{align}
\label{eq:vector}
    \mathcal{D}_{\mu} \mathcal{N} = (\partial_{\mu} + \Gamma_{\mu} - i v_{\mu}^{(s)}) \mathcal{N},
\end{align}
where $\mathcal{N} = \begin{pmatrix} p \\n \end{pmatrix}$ is the nucleon isodoublet ($p$ is a proton and $n$ is a neutron) and $v_{\mu}^{(s)} = - e A_{\mu} \mathbb{I} / 2$. The axial vector field $u_\mu$ couples to the nucleon doublet through the Lagrangian term
\begin{align}
\label{eq:axial}
    \mathcal{L} \supset \frac{g_A}{2} \bar{\mathcal{N}}\gamma^{\mu} \gamma^5 u_{\mu} \mathcal{N}, 
\end{align}
where the axial coupling $g_A = 1.27$ is determined from neutron beta decay~\cite{Czarnecki:2018okw}. The nucleon covariant derivative yields pion-nucleon vertices with even numbers of pions while the axial term contains vertices with odd numbers of pions. 

Combining the pion and nucleon terms, our Lagrangian including nuclear and gauge interactions to lowest order in derivatives is
\begin{equation}
\begin{split}
    \mathcal{L} = \frac{f_{\pi}^2}{4} \text{Tr}( D_{\mu} U^{\dagger} D^{\mu} U) + \frac{1}{4} m_{\pi}^2 f_{\pi}^2 \text{Tr}(U + U^{\dagger} - 2) \\
    + \bar{\mathcal{N}} \left(i \gamma^{\mu} \mathcal{D}_{\mu} + \frac{g_A}{2} \gamma^{\mu} \gamma^5 u_{\mu} -M \right)\mathcal{N},
    \label{eq:lagrangian}
\end{split}
\end{equation}
where $M$ is the nucleon mass. By performing the field redefinition in Eq.~(\ref{eq:massbasis}), we obtain the dark photon couplings to pions and nucleons by replacing the photon field $A_\mu$ with $-\epsilon A'_\mu$.

The complete Lagrangian terms with pion-nucleon interactions are
\begin{equation}
\label{eq:lagpN}
    \begin{split}
        \mathcal{L}_{\pi N} \supset \frac{g_A}{2 f_{\pi}} \Big(\bar{p} \gamma^{\mu} \gamma^5 p \, \partial_{\mu} \pi^0 - \bar{n} \gamma^{\mu} \gamma^5 n\, \partial_{\mu} \pi^0   \\
        +\sqrt{2} \, \bar{p} \gamma^{\mu} \gamma^5 n \, \partial_{\mu} \pi^+ + \sqrt{2} \, \bar{n} \gamma^{\mu} \gamma^5 p \, \partial_{\mu} \pi^- \Big) \\
        +\frac{i}{4 f_\pi^2} \Big( \bar{p} \gamma^\mu p \, (\pi^+ \partial_\mu \pi^- - \pi^- \partial_\mu \pi^+) \\
        + \bar{n} \gamma^\mu n \, (\pi^- \partial_\mu \pi^+ - \pi^+ \partial_\mu \pi^-) \\
        + \sqrt{2} \, \bar{n} \gamma^\mu p (\pi^- \partial_\mu \pi^0 - \pi^0 \partial_\mu \pi^-)\\
        + \sqrt{2} \, \bar{p} \gamma^\mu n (\pi^0 \partial_\mu \pi^+ - \pi^+ \partial_\mu \pi^0) \Big),
    \end{split}
\end{equation}
while the Lagrangian terms with pion-nucleon-dark photon interactions are
\begin{equation}
\label{eq:lagpApr}
    \begin{split}
        \mathcal{L}_{\pi N A'} \supset i e \epsilon A' (\pi^- \partial_\mu \pi^+ - \pi^+ \partial_\mu \pi^-) \\
        + e^2 \epsilon^2 A'_\mu A'^\mu \, \pi^+ \pi^- 
        + e \epsilon \, A'_\mu \, \bar{p} \gamma^\mu p 
        \\
        + \frac{i e \epsilon g_A}{\sqrt{2} f_{\pi}}  
         A'_{\mu} \Big(\bar{n} \gamma^{\mu} \gamma^5 p \, \pi^- - \bar{p} \gamma^\mu \gamma^5 n \, \pi^+\Big)  \\
         + \frac{e \epsilon}{2 f_\pi^2} A'_\mu \Big(\bar{n} \gamma^\mu n \, \pi^+ \pi^- 
         -\bar{p} \gamma^\mu p \, \pi^+ \pi^- \\
         + \frac{1}{\sqrt{2}}\, \bar{n} \gamma^\mu p\, \pi^0 \pi^- +\frac{1}{\sqrt{2}}\, \bar{p} \gamma^\mu n \, \pi^0 \pi^+ \Big).
    \end{split}
\end{equation}

The leading-order Feynman diagrams which yield dark photon bremsstrahlung for pion-proton scattering are shown in Fig.~\ref{fig:ChiralFeynmanDiags}, with an analogous set of diagrams appropriate for neutron scattering. Since the $A'$ couples to the conserved QED current of pions and protons, we generate the expected 3-point $\pi \pi A'$ and $\mathcal{N} \mathcal{N} A'$ interactions as well as the $\pi \mathcal{N} \mathcal{N} A'$ 4-point and $\pi \pi \mathcal{N} \mathcal{N} A'$ 5-point contact terms.
We checked that the Ward identity was satisfied for these amplitudes; this ensures that we are using the complete set of diagrams for our scattering process. We neglect the $\Delta$ resonance as well as low-lying vector resonances like the $\rho$, whose contributions we will study in future work and which will likely further increase the $A'$ production rate.

\subsubsection{Kinematics of nuclear scattering}

For fixed-target scattering experiments, the nucleons are typically embedded in a large nucleus, and thus connecting the nucleon-level chiral perturbation theory calculation to the kinematics of scattering off a heavy nucleus requires some assumptions about the distribution of nucleons inside the nucleus. While there are many nuclear theory models for these distributions, which are crucial for an accurate treatment of e.g.\ neutrino-nucleus scattering~\cite{SajjadAthar:2022pjt}, the primary effect in our setup is purely kinematical: the maximum energy taken by the dark photon depends on the physical mass of the target $M_T$. Specifically, as shown in Appendix~\ref{app:CS}, the dark photon energy fraction $x \equiv E_{A'}/E_\pi$ is at most
\begin{equation}
\label{eq:xmax}
    x_{\rm max} = \frac{2 E_{\pi} M_T - m_{A'}^2 - 2 M_T m_{\pi}}{2 E_{\pi}(M_T + m_\pi)}.
\end{equation} 
For example, with $m_{A'} = 1 \ {\rm GeV}$ and $E_\pi = 30 \ {\rm GeV}$, $x_{\max} = 0.85$ for $M_T = M$, the nucleon mass (these kinematics correspond to the impulse approximation of quasi-elastic nuclear scattering), but $x_{\rm max} = 0.99$ for $M_T = 56 M$ as would be the case for an iron nucleus. 

As alluded to in the Introduction and as we will demonstrate further below, the matrix element for $m_{A'} > m_\pi$ peaks at $x \approx 1$, while the overall magnitude of the cross section is largely independent of $M_T$. Thus, accessing the dominant region of phase space would require that some nucleons be off-shell and effectively carry most of the mass of the nucleus. Rather than attempting to model this effect directly, in what follows we will present our results for nucleons of mass $M_T$ varying between $M$ and $M_N$ (the latter being the full mass of the nucleus) in order to illustrate the size of the uncertainty.\footnote{Note  that no such uncertainties would arise in directly using the nuclear quark PDFs to compute parton-level scattering rates.}

\subsubsection{Pion-nucleus scattering}

\begin{figure*}[t]
    \centering
    \includegraphics[width = 0.31\textwidth]{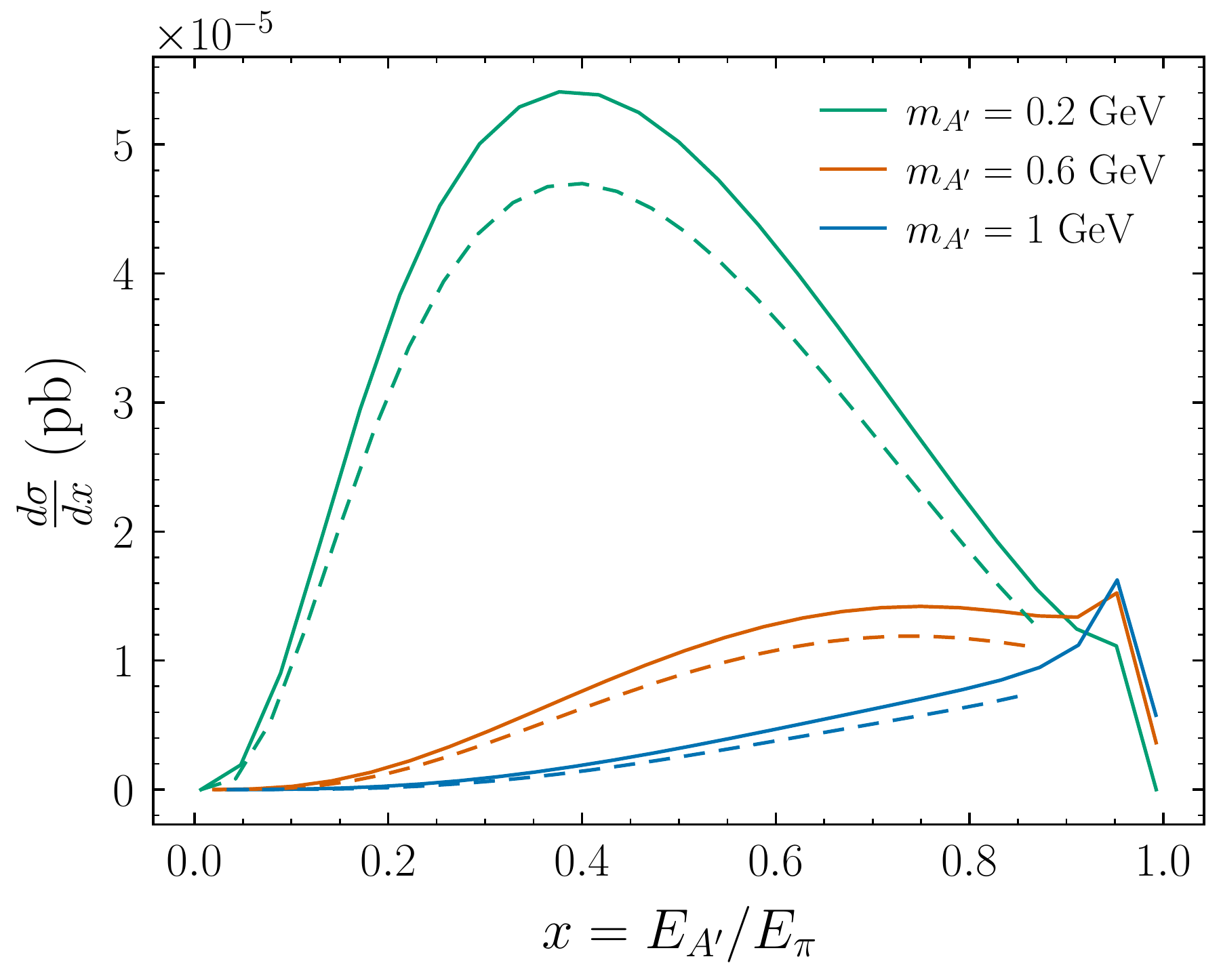}
    \hspace{0.2cm}
    \includegraphics[width = 0.31\textwidth]{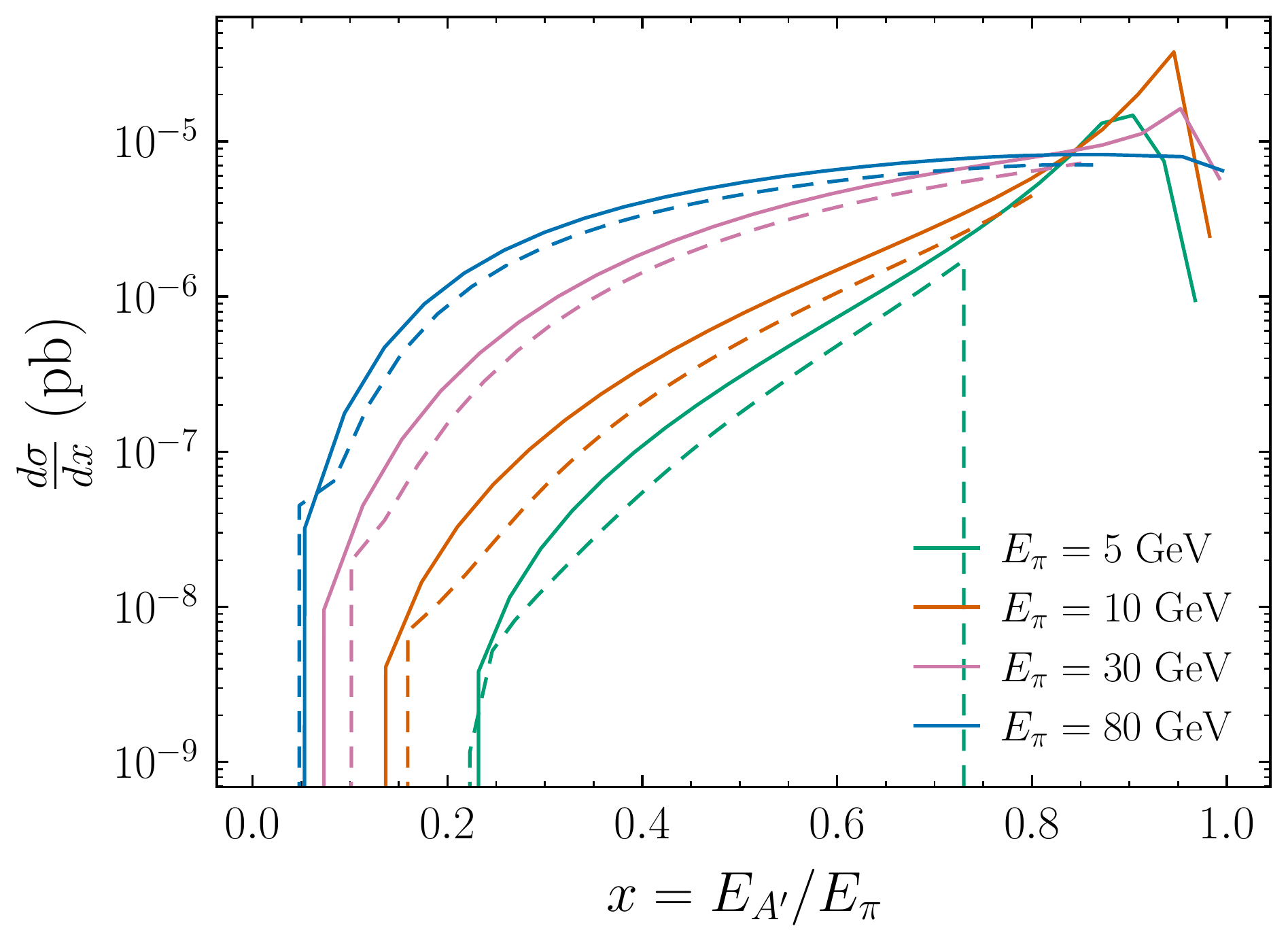}
    \hspace{0.2cm}
    \includegraphics[width = 0.31\textwidth]{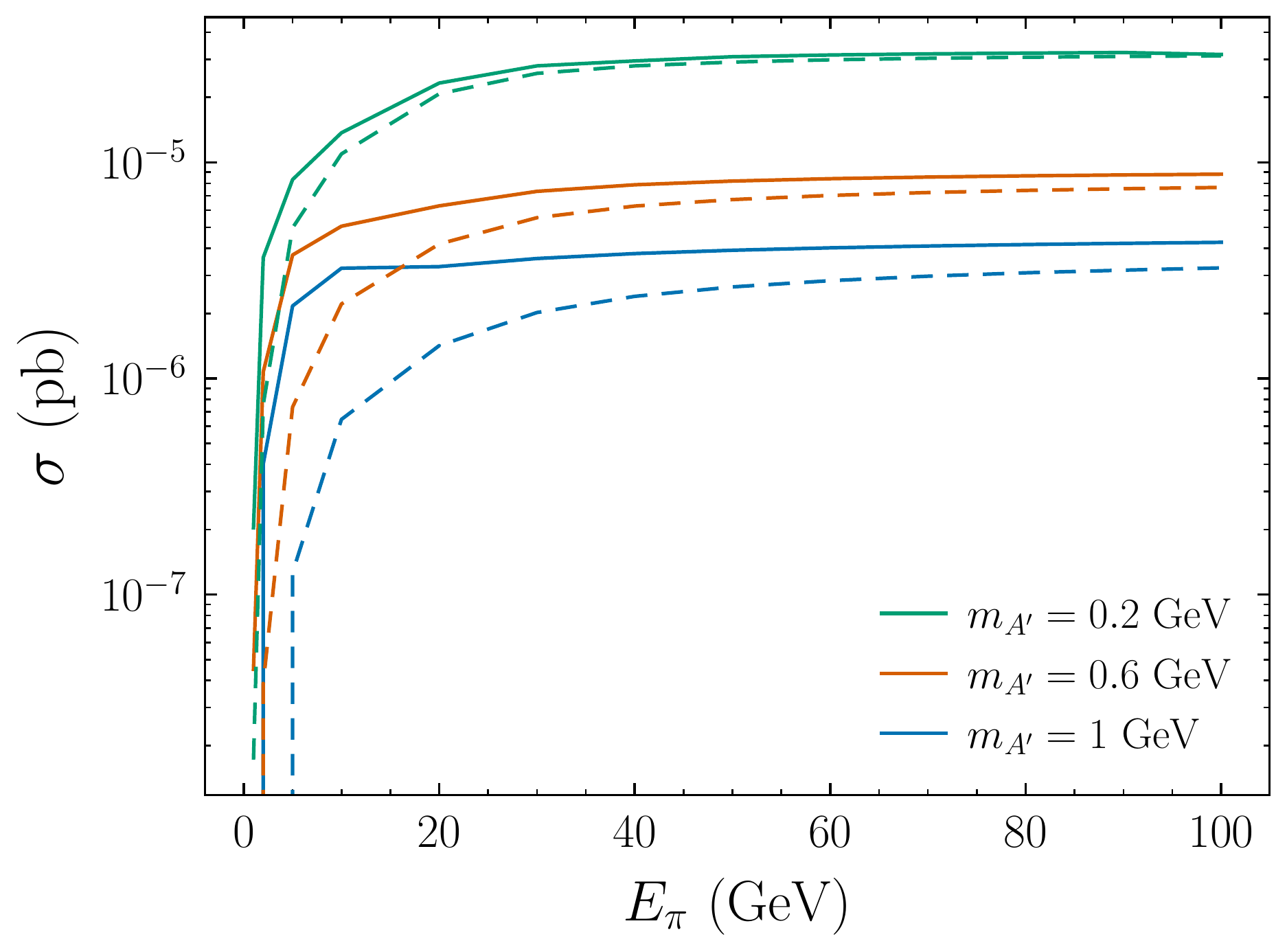}
    \caption{\textbf{Left:} Differential cross section for the energy fraction taken by the $A'$ from a pion beam of energy $E_{\pi} = 30$ GeV. As in the case of emission from lepton beams, the distribution peals at $x = 1$ for $m_{A'} \gg m_\pi$. Dashed lines correspond to $M_T = M$, and solid lines correspond to $M_T = M_N$. Apart from the kinematic cutoff due to the different $x_{\rm max}$, the effect of varying the target mass is only at the 20\% level. \textbf{Center:} Differential cross sections for $A'$ production with $m_{A'} = 1$ GeV and $\epsilon = 10^{-7}$, for various pion beam energies. Each $E_{\pi}$ has a different kinematically allowed $A'$ energy range, as shown by the differing minimum/maximum values of $x$, with differences between the dashed ($M_T = M$) and solid ($M_T = M_N$) curves most apparent near $x_{\rm max}$. The shape of the curves is also dictated by the chosen angular acceptance. The suppression of the cross section at lower $x$ is due to restricting the emission angle to $\theta <  0.05$ radians as determined by the experimental geometry. See Sec.~\ref{sec:signals} for more details on the experimental acceptance. \textbf{Right:} Total cross section as a function of the incoming pion beam energy for different $A'$ masses and $\epsilon = 10^{-7}$. Well above threshold, the cross section is relatively insensitive to $E_\pi$. The differences between the solid and dashed lines illustrate the effect of varying $M_T$ between $M$ and $M_N$, with the larger cross sections corresponding to $M_T = M_N$ from integrating over the region near $x = 1$.}
    \label{fig:dsigmadx}
\end{figure*} 

Following Ref.~\cite{Liu:2016mqv}, we computed the cross section for $\pi N \to \pi N A'$ in a form suitable for deterministic numerical integration, which facilitates calculating differential cross sections. The total $2 \to 3$ cross section for an incoming pion with 4-momentum $(E_\pi, \pp_\pi)$ scattering from a nucleus of atomic number $Z$ and mass number $A$ is modeled as incoherent quasi-elastic scattering off each individual nucleon of mass $M_T$, weighting the proton cross section by the $Z$ and the neutron cross section by $A - Z$. The cross section is given in terms of a 3-body phase space integral, 
\begin{equation}
\begin{split}
    \sigma & = \frac{1}{1024 \pi^4 M_T^2} \int_{t_{\rm min}}^{t_{\rm max}}   dt  \int_{\cos \theta_{\rm max}}^1 d \cos \theta \int_0^{2\pi} d\phi_q  
     \\
    & \times \int_{x_{\rm min}}^{x_{\rm max}} dx \frac{|\kk|}{ |\V|} \, \left(Z\langle|\mathcal{M}_{2 \rightarrow 3}^{(p)}|^2\rangle + (A-Z)\langle|\mathcal{M}_{2 \rightarrow 3}^{(n)}|^2\rangle \right),
    \label{eq:totalcs}
\end{split}
\end{equation}
where the dark photon 3-momentum is $\kk = (\sqrt{(x E_{\pi})^2 - \mA^2} \cos{\theta},0,\sqrt{(x E_{\pi})^2 - \mA^2} \sin{\theta})$, $\V = \kk - \pp_{\pi}$, $\langle |\mathcal{M}_{2 \rightarrow 3}^{(p,n)}|^2 \rangle$ is the spin-averaged squared matrix element for $\pi^{\pm} p (n) \to \pi^{\pm} p (n) A'$\footnote{The chiral Lagrangian also allows isospin-exchange processes, such as $\pi^+ \, n \to \pi^0 \, p \, A'$, which we neglect for simplicity; such processes would only increase the total $A'$ production rate.}, and the integration variables are the squared momentum transfer to the nucleus $t$, the azimuthal angle of the momentum transfer $\phi_q$, the dark photon emission angle $\theta$, and the dark photon energy fraction  $x = E_{A'} / E_{\pi}$.  Note that unlike dark photon processes initiated by electromagnetic scattering, it is not obvious that a Weizs\"{a}cker-Williams approximation \cite{Liu:2016mqv,Liu:2017htz} can be used to reduce the cross section to a simpler $2 \to 2$ process, since the scattering occurs entirely through contact interactions or exchanges of massive nucleons. 

As mentioned above, since we are working in the regime of chiral perturbation theory, we cut off the $t$ integral at $t_{\rm max} = (4 \pi f_\pi)^2$. As shown in App.~\ref{app:CS}, $\cos \theta_{\rm max}$ is determined by the experimental geometry, the limits on $x$ only depend on $E_\pi$ and $m_{A'}$, and $t_{\rm min}$ depends on both $x$ and $\theta$. In practice, imposing $t \leq (4\pi f_\pi)^2$ further restricts the emission angle to the forward direction. 

Fig.~\ref{fig:dsigmadx} (left) shows the energy fraction distribution of the $A'$ produced in charged pion bremsstrahlung, with a broad peak at $x \simeq 0.5$ for $m_{A'} \simeq m_\pi$ but with a sharper peak at $x \approx 1$ for $m_{A'} \gg m_\pi$. This is the same phenomenology previously observed in massive dark photon emission from lepton beams. Fig.~\ref{fig:dsigmadx} (center) plots the cross section as a function of $x$ for fixed $m_{A'}$ but various pion energies $E_\pi$, showing the change in $x_{\rm min}$ at threshold and the suppression of the cross section at small $x$ due to the restriction of the angular integral to the experimental geometry. Finally, Fig.~\ref{fig:dsigmadx} (right) shows the total cross section for $A'$ production as a function of pion energy. Above threshold, the cross section is quite flat and relatively insensitive to $m_{A'}$. In all cases the effect of changing the target mass is at the 20\% level except near the kinematic boundary $x_{\rm max}$, which can affect the total cross section at the $\mathcal{O}(1)$ level.

We can understand the order of magnitude of the cross section in Fig.~\ref{fig:dsigmadx} from a dimensional analysis argument, as follows. From the Lagrangian in Eqs.~(\ref{eq:lagpN})--(\ref{eq:lagpApr}), each diagram scales as $\epsilon e/f_\pi^2$. The squared matrix element picks up factors of $M_T$ from the external nucleon spinors, and with $\mathcal{O}(10)$ diagrams adding incoherently, summing incoherently over the $A = 56$ nucleons in an iron nucleus gives $|M_{2 \to 3}|^2 \sim 600 \epsilon^2 e^2 M_T^2/f_\pi^4$. Since the cross section peaks sharply at $\cos \theta = 1$ (see Fig.~\ref{fig:thetadistribution} in App.~\ref{app:CS}), we can approximate the matrix element as being proportional to a delta function $\delta(\cos \theta - 1)$, such that integrating over phase space with $t_{\rm max} = (4\pi f_\pi)^2$ gives
\begin{align}
    \sigma_{\rm Fe} & \sim \frac{600 \epsilon^2 e^2 M_T^2/f_\pi^4}{1024 \pi^4 M_T^2}(2\pi)(4\pi f_\pi)^2 \\
    & \sim \frac{75 \alpha \epsilon^2}{f_\pi^2} = 2.4 \times 10^{-4} \ {\rm pb} \left(\frac{\epsilon}{10^{-7}}\right)^2.
    \label{eq:csparametrics}
\end{align}
Comparing to Fig.~\ref{fig:dsigmadx} (right), this overestimates the cross section by one or two orders of magnitude, due to the combinatorical factors and interference terms in the many diagrams as well as the kinematic boundaries for larger $m_{A'}$. Regardless, comparing to the $A'$ cross section from inelastic proton bremsstrahlung~\cite{Berlin:2018pwi},
\begin{equation}
\label{eq:pbremparametrics}
\sigma_{p-{\rm brem}} \sim A \alpha \epsilon^2 \sigma_{pp} \sim 2.0 \times 10^{-4} \ {\rm pb} \left(\frac{\epsilon}{10^{-7}}\right)^2,
\end{equation}
we see that pion bremsstrahlung production is parametrically of the same order (indeed, the inelastic proton cross section $\sigma_{pp}$ is essentially the same as the pion cross section $1/f_\pi^2$) but can leverage the large multiplicity of secondary pions.

\subsection{Dark photon flux}

%
\begin{figure}
    \centering
    \includegraphics[width = 0.47\textwidth]{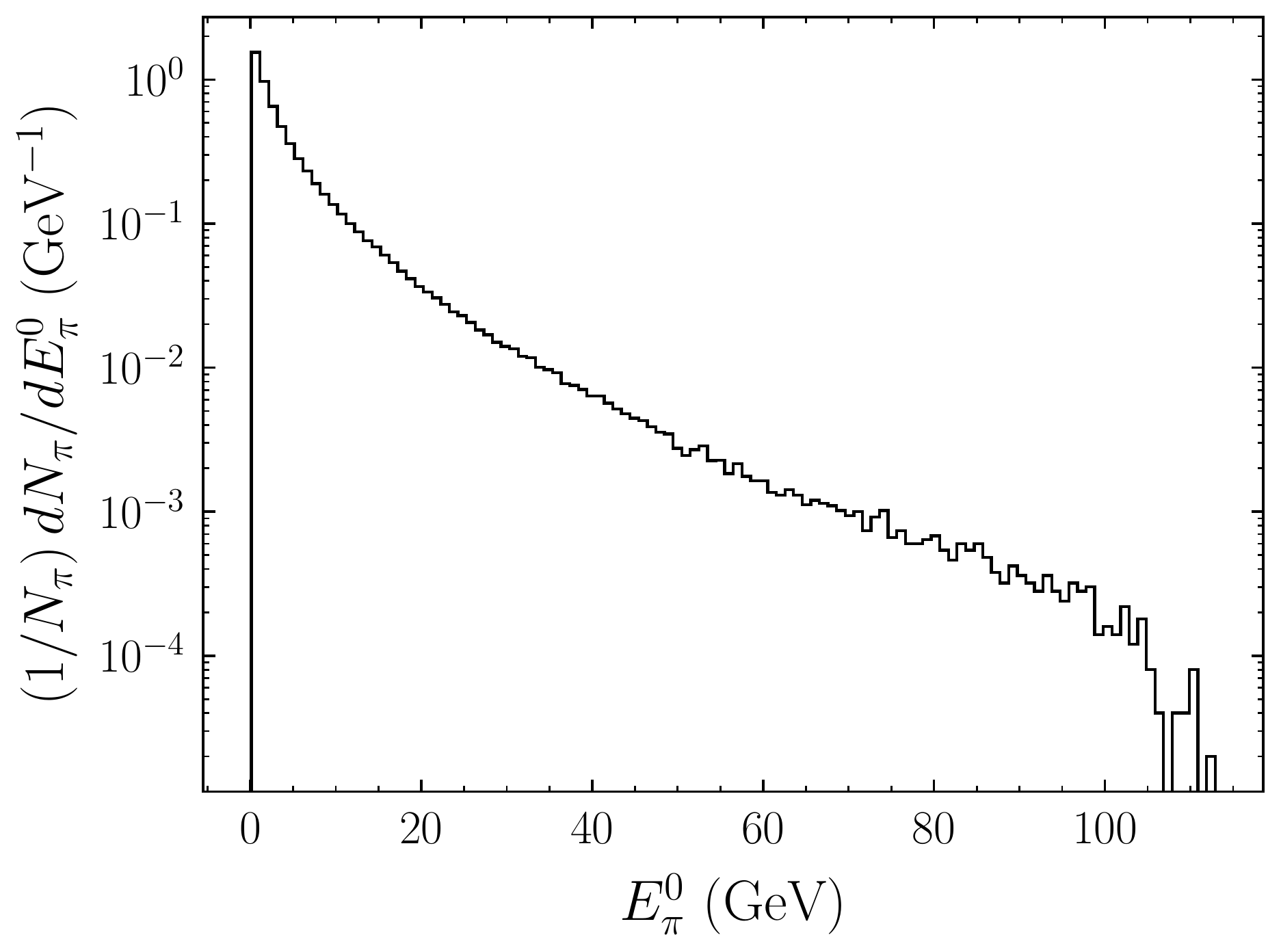}
    \caption{The energy distribution of charged pions from a \texttt{Pythia 8.2} simulation of a 120 GeV proton beam on an iron dump. The pions are created in the first thin slice of the iron dump. The distribution was adapted from the analysis of \cite{Berlin:2018pwi}.
    }
    \label{fig:piondist}
\end{figure}

Using (\ref{eq:totalcs}), we can calculate the total flux dark photons produced through charged pion bremsstrahlung. The number of dark photons is given by 
\begin{equation}
    N_{A'} =  n_T L (N_{\pi^+}\sigma_{\pi^+} + N_{\pi^-}\sigma_{\pi^-}),
\end{equation}
where $n_T$ is the density of the target material, $L$ is the total length of the target, $N_{\pi^\pm}$ is the number of positive or negative secondary pions produced from the primary proton beam, and $\sigma_{\pi^\pm}$ is the cross section for $A'$ production from the correspondingly-charged pion as discussed in Sec.~\ref{sec:cs}. For a given $m_{A'}$ and $\epsilon$, the cross section also depends on the dark photon's energy, $E_{A'}$, and the pion's energy, $E_{\pi}(l)$, which is a function of the position $l$ in the target. As the pion scatters through the material, it loses energy as $E_{\pi}(l) = E_{\pi}^0 \exp(-l / l_{\pi})$, where $E_{\pi}^0$ is the initial pion energy and $l_{\pi}$ is the pion interaction length. Following Ref.~\cite{Berlin:2018pwi}, we use a sample of pions generated in a a \texttt{Pythia 8.2} simulation using a 120 GeV proton beam incident on an iron target, from which the distribution of $E_\pi^0$ is shown in Fig.~\ref{fig:piondist}. For an iron target, $l_{\pi} \sim 20$ cm \cite{ParticleDataGroup:2016lqr}. The total pion multiplicity is $N_{\pi^+} + N_{\pi^-} \sim 6.5 \ {\rm POT}$.

We can thus write the number of dark photons produced as
\begin{equation}
    N_{A'} = n_T \int_0^L dl \int_{x_{\rm min}}^{x_{\rm max}} dx \, \left( N_{\pi^+} \, \frac{d \sigma_{\pi^+}}{d x} +  N_{\pi^-} \, \frac{d \sigma_{\pi^-}}{d x}\right),
    \label{eq:Aproduction}
\end{equation}
where we changed variables to $x = \frac{E_{A'}}{E_{\pi}(l)}$. Here, for a fixed $m_{A'}$ and $\epsilon$, the differential cross section $d\sigma / dx$ is only a function of $x$ and $E_{\pi}(l)$ (and is in general slightly different for $\pi^+$ and $\pi^-$). 
Here we have implicitly
integrated over the $A'$ emission angle $\theta$ corresponding to the acceptance of the beam dump experiment's geometry; 
see Sec. \ref{sec:signals} for more details. 

\begin{figure}
    \centering
    \includegraphics[width=0.47\textwidth]{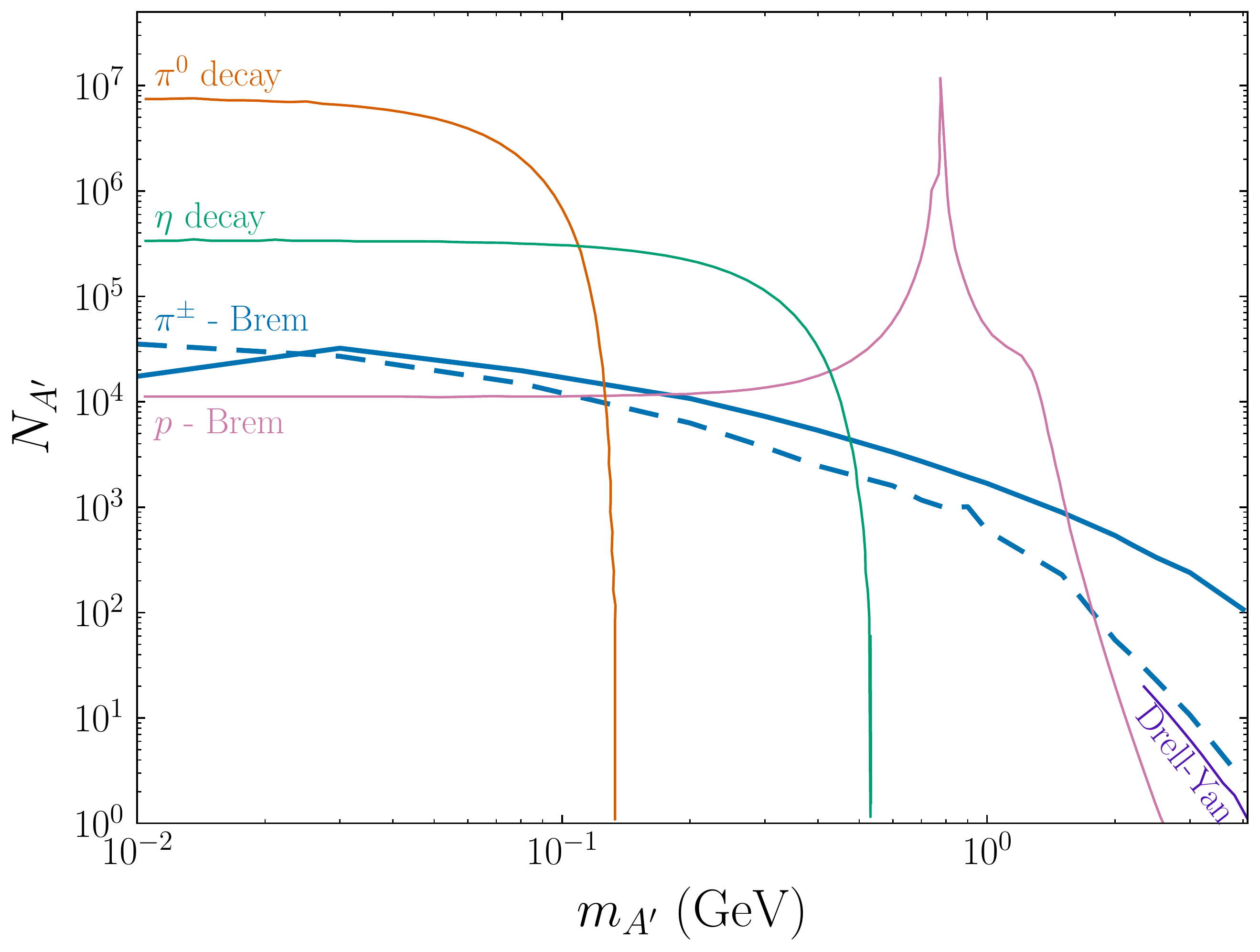}
    \caption{The number of dark photons expected at SpinQuest with $1.44 \times 10^{18}$ POT for $\epsilon = 10^{-6}$. We compare dark photon production from charged pion bremsstrahlung (blue, solid line has $M_T = M_N$ and dashed line has $M_T = M$) to production from pseudoscalar meson decay, $\pi^0 \rightarrow \gamma A'$ (orange) and $\eta \rightarrow \gamma A'$ (green), as well as proton bremsstrahlung (pink) and Drell-Yan production (purple). The curves for meson decay, proton bremsstrahlung, and Drell-Yan production were adapted from Ref.~\cite{Berlin:2018pwi}. }
    \label{fig:NAproduced_ep1e-6}
\end{figure}

Figure \ref{fig:NAproduced_ep1e-6} shows the number of $A'$s produced from secondary charged pion bremsstrahlung in iron ($Z = 26$, $A = 56$) compared to the previously-considered channels of $\pi^0$ and $\eta$ decay, as well as primary proton bremsstrahlung. As anticipated in Eqs.~(\ref{eq:csparametrics})--(\ref{eq:pbremparametrics}), production from charged pion bremsstrahlung is comparable to proton bremsstrahlung except for the region around the $\rho$ resonance (which we have neglected in our analysis), and dominates over proton bremsstrahlung at low dark photon masses due to the pion multiplicity. Even with the momentum cutoff at $4 \pi f_\pi$, secondary dark photon production falls off much less rapidly at high masses than either of the production processes from the primary proton beam, making this channel especially useful for dark photon searches above 1.5 GeV where it is the dominant process even accounting for nuclear modeling uncertainties. For $m_{A'} < m_\pi$, charged pion bremsstrahlung suffers a 3-body phase space penalty compared to the decays $\pi^0, \eta \to \gamma A'$. Moreover, at these low masses the kinematics of charged pion bremsstrahlung approaches that of ordinary QED bremsstrahlung, with $d\sigma/dx$ peaking near $x = 0$, as opposed to the highly-boosted $A'$s from $\pi^0$ decay.

\subsection{Dark photon decays}
Once a dark photon is produced, it may be detected through its decay into charged leptons or hadrons \cite{Bauer:2018onh}. The decay rate to charged leptons for $m_l \ll m_{A'}$ is 
\begin{equation}
    \Gamma(A' \rightarrow l^+ l^-) = \frac{\alpha \epsilon^2 m_{A'}}{3}.
    \label{eq:leptondecay}
\end{equation}
For the decay rate to hadrons, we use 
\begin{equation}
    \Gamma(A' \rightarrow \text{hadrons}) = \Gamma(A' \rightarrow \mu^+ \mu^-) \, R(m_{A'}^2),
    \label{eq:hadrondecay}
\end{equation}
where $R(s) = \sigma(e^+ e^- \rightarrow \text{hadrons})/ \sigma(e^+ e^- \rightarrow \mu^+ \mu^-)$, see~\cite{ParticleDataGroup:2016lqr}. For dark photon masses $m_{A'} \lesssim 0.6 $ GeV, electrons and muons are the majority of the decay products, while for larger masses, hadrons are the dominant decay channel~\cite{Bauer:2018onh}. 

To be detected at a beam dump experiment, the dark photon must decay within the fiducial decay region of the detector.  Eqs.~(\ref{eq:leptondecay}) and~(\ref{eq:hadrondecay}) yield the lab-frame decay length for a dark photon:
\begin{equation}
    \begin{split}
    L_{A'} \equiv \gamma c \tau_{A'} \approx 10 \, \text{m} \left(\frac{E_{A'}}{5 \text{ GeV}} \right) \left(\frac{1 \ \text{GeV}}{m_{A'}} \right) \left(\frac{10^{-7}}{\epsilon} \right)^2.
    \end{split}
\end{equation}
From this estimate, combined with the flux of dark photons from Fig.~\ref{fig:NAproduced_ep1e-6}, we can see that dark photons with $\epsilon \sim 10^{-7}$ produced from GeV-energy pions could reasonably decay within the length of a beam dump experiment such as SpinQuest. For long-lived particle detectors at the LHC which are $\mathcal{O}( 100 \ {\rm m})$ from the interaction point, probing dark photons at this mass and coupling requires more energetic pions and suffers a corresponding flux penalty.

At a given experiment with sensitivity to a final-state particle $X$, the number of signal events given by dark photons decaying to $X$ is
\begin{equation}
\begin{split}
     N_{\rm sig} = N_{A'} \times \, \text{BR}(A' \rightarrow XX) \, \times P_{\rm decay} 
     \label{eq:Nsignal}
\end{split}
\end{equation}
where $P_{\rm decay}$ is the probability that $A'$ decays within the fiducial decay region. In the following section, we will define $P_{\rm decay}$ according to specific experimental geometry and use (\ref{eq:Nsignal}) to determine the sensitivity of the SpinQuest experiment at Fermilab to the additional source of dark photons we have described.

\section{Signal Events at SpinQuest/DarkQuest}
\label{sec:signals}

\subsection{SpinQuest/DarkQuest Setup and Signal Acceptance}

The SpinQuest spectrometer was originally designed to study di-muon production in Drell-Yan processes from the proton beam collisions on nuclear targets~\cite{Brown:2014sea}. Currently, SpinQuest is operating at Fermilab with a 120 GeV proton beam on an iron target, and is expected to achieve an integrated luminosity of $1.44 \times 10^{18}$ POT by the end of 2024. The SpinQuest spectrometer spans $\sim 25$ m. The first 5 m is comprised of a magnetized iron dump (``FMAG"). The FMAG's magnetic field imparts a kick of $\Delta p_T \simeq 2.9$ GeV to sweep away soft particles. Additionally, 9 m down the beam line, a 3 m magnet (``KMAG") is placed between the first two tracking detectors. The KMAG imparts a kick of $\Delta p_T \simeq 0.4$ GeV to further remove soft particles and aid in momentum reconstruction. An absorber 20 m down the beamline stops all SM particles other than muons, which pass to a muon ID station; three tracking stations are also placed between the magnets and the absorber in order to accurately reconstruct the muon momentum. The proposed DarkQuest upgrade will add an electromagnetic calorimeter (ECAL) after tracking station 3, in order to identify electrons, photons, and charged pions and expand the search for dark sector physics \cite{Apyan:2022tsd}. In this analysis, we focus on decays to leptons only, though including the channel $A' \to \pi^+ \pi^-$ may increase the sensitivity considerably for heavy $A'$.\footnote{There have been some preliminary studies of $e/\pi$ discrimination at DarkQuest~\cite{Apyan:2022tsd}, but for our purposes, no such discrimination is required so long as the invariant mass can be reconstructed because our signal is an inclusive decay rate. We thank Nhan Tran for clarification on this point.}

To model DarkQuest's experimental acceptance, we will follow the analysis laid out in \cite{Berlin:2018pwi}. For $A'$ decays to electrons or muons, we only include events in which the leptons are captured by station 3 and detected by at least one other tracking station. The dark photon can then decay in two fiducial decay regions: 
\begin{enumerate}
    \item 5--6 m: After FMAG and before station 1
    \item 5--12 m: After FMAG and before end of KMAG 
\end{enumerate}
The location of the $A'$ decay determines how well the lepton decay products can be tracked in DarkQuest.
The momentum and vertex of the leptons created in Region 1 can be accurately reconstructed since the leptons travel through all tracking stations. Therefore, estimating the sensitivty based only on the 5--6 m fiducial decay region is the most conservative, and is currently the basis for the analysis strategy in SpinQuest. For electrons created in Region 2, the ECAL can determine the electrons' energies which partially mitigates the reduced tracking and vertexing capabilities. With sufficiently low backgrounds, the ECAL allows us to expand our analysis to the larger fiducial decay region, increasing the sensitivity. 

To determine if the dark photon decays within the fiducial decay region and its decay products meet DarkQuest's geometric acceptance, we define the probability of decay as in \cite{Berlin:2018pwi}:
\begin{equation}
    P_{\rm decay} = \frac{1}{N_{\rm MC}} \sum_{\rm events \,  \in \,  geom.}  \int_{z_{\rm min}}^{z_{\rm max}} dz\,   \Gamma \, \frac{e^{-z (m / p_z) \Gamma}}{p_z/m},
    \label{eq:pdecay}
\end{equation}
where $z_{\rm min}$ and $z_{\rm max}$ are the boundaries of the fiducial decay regions, $N_{\rm MC}$ is the number of simulated Monte Carlo decay events, and $m$, $\Gamma$, and $p_z$ are the mass, decay width, and momentum of the dark photon. 
$P_{\rm decay}$ factorizes into 
\begin{equation}
    P_{\rm decay} \propto \mathcal{G} \times  e^{-l_{\rm tar} / (\gamma c \tau_{A'})} \times (1-e^{-l_{\rm dec} / (\gamma c \tau_{A'})}),
\end{equation}
which is the decay probability normally cited in the literature with an added factor of $\mathcal{G}$ to account for the experiment's geometry. However, (\ref{eq:pdecay}) automatically takes into account the geometric acceptance of the SpinQuest experiment. Since we only accept signal events that reach station 3, we only consider the dark photons that decay within angles $\sin \theta \lesssim 0.05$. 

The main background for $A' \to e^+ e^-$ comes from semileptonic $K_L^0$ decays, $K_L^0 \to \pi^\pm \ell^\mp \nu_\ell$, with the pion faking an electron. This background is small enough to neglect for the 5--6 m fiducial decay region, but may be significant for the 5--12 m region \cite{Berlin:2018pwi}. The main backgrounds for $\mu^+ \mu$ are likely combinatoric from secondary muons produced in the dump, with a rate which is difficult to estimate without dedicated studies~\cite{TranPrivate}. Following Ref.~\cite{Berlin:2018pwi} in order to compare with previous analyses, we will assume that the signal-to-background is sufficiently large for a signal rate of 10 events, and just consider the most conservative decay region for the current SpinQuest operations, with $1.44 \times 10^{18}$ protons on target (POT). For future projections, where DarkQuest is proposed to have $10^{20}$ POT, we will assume the same 10-event sensitivity.

\subsection{Monte Carlo Setup}

In order to estimate the sensitivity of SpinQuest/DarkQuest to dark photons, we use a Monte Carlo simulation to calculate (\ref{eq:Aproduction}), (\ref{eq:Nsignal}), and (\ref{eq:pdecay}). The simulation consists of three parts: (1) the generation of secondary charged pions; (2) the production of dark photons in the FMAG, accounting for the energy loss of pions as they traverse the dump; and (3) the decay of dark photons in the fiducial decay region(s).  

The total number of secondary charged pions can be written as an integral over the energy spectrum of charged pions at the instant the pions are produced:
\begin{equation}
    N_{\pi} = \int d E^0_{\pi} \frac{dN_\pi}{d E_{\pi}^0}.
\end{equation}
To simplify the analysis, we assume that all secondary particles were created at the front end of KMAG, closest to the beam source, and that isospin symmetry holds, $N_{\pi^+} = N_{\pi^-} = N_\pi/2$.\footnote{We find that this is also a good approximation for the chiral perturbation theory cross sections; $\sigma_{\pi^+} \approx \sigma_{\pi^-}$ for iron.} Fig.~\ref{fig:piondist} shows that the majority of pions are low-energy compared to the beam, $E_{\pi}^0\sim 1-10 $ GeV. However, as previously emphasized, the multiplicity of the secondaries is large, $N_{\pi} \sim 6.5 \ {\rm POT}$.

\begin{figure*}[t]
    \centering
    \includegraphics[width=0.47\textwidth]{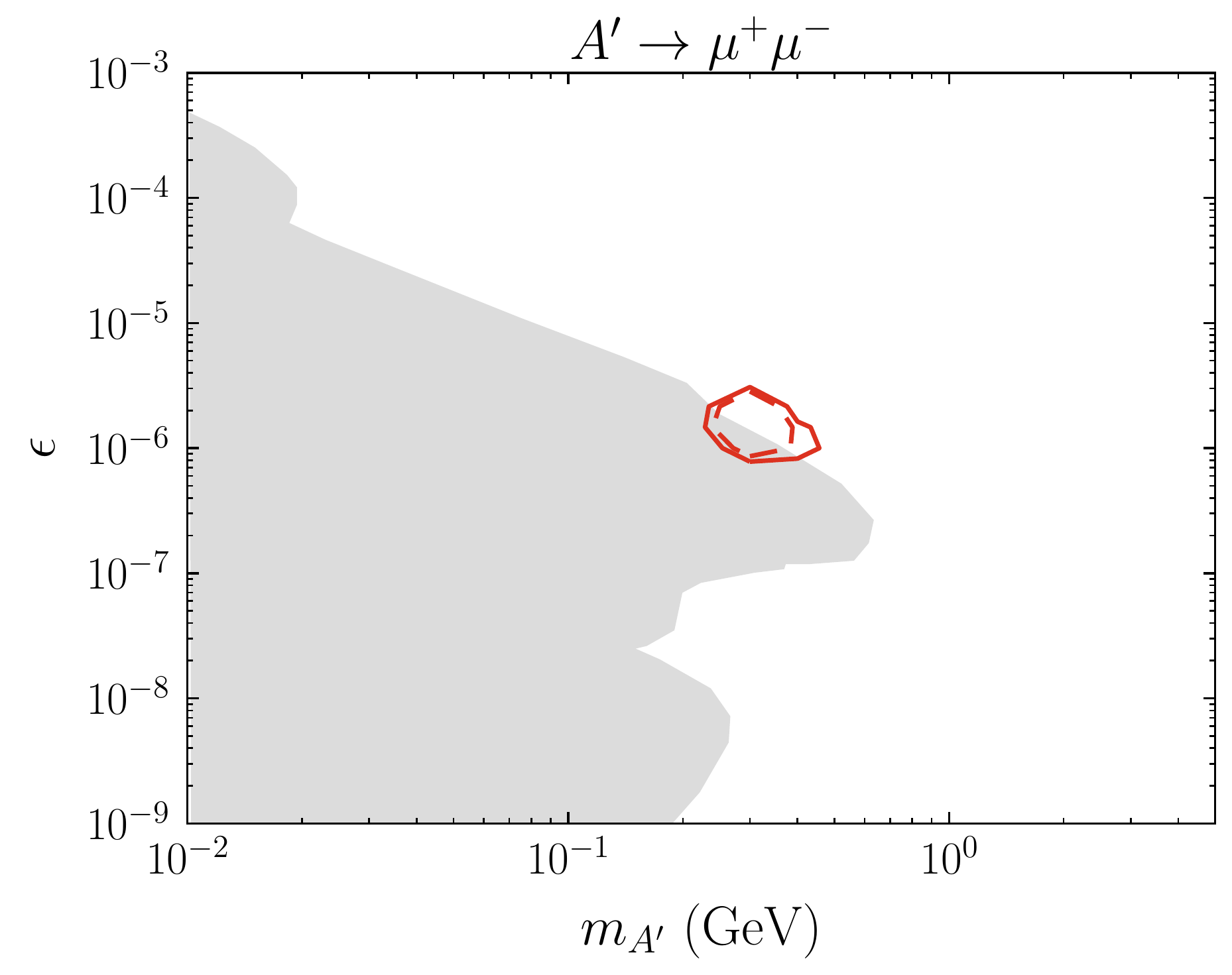}
    \hspace{0.2cm}
    \includegraphics[width=0.47\textwidth]{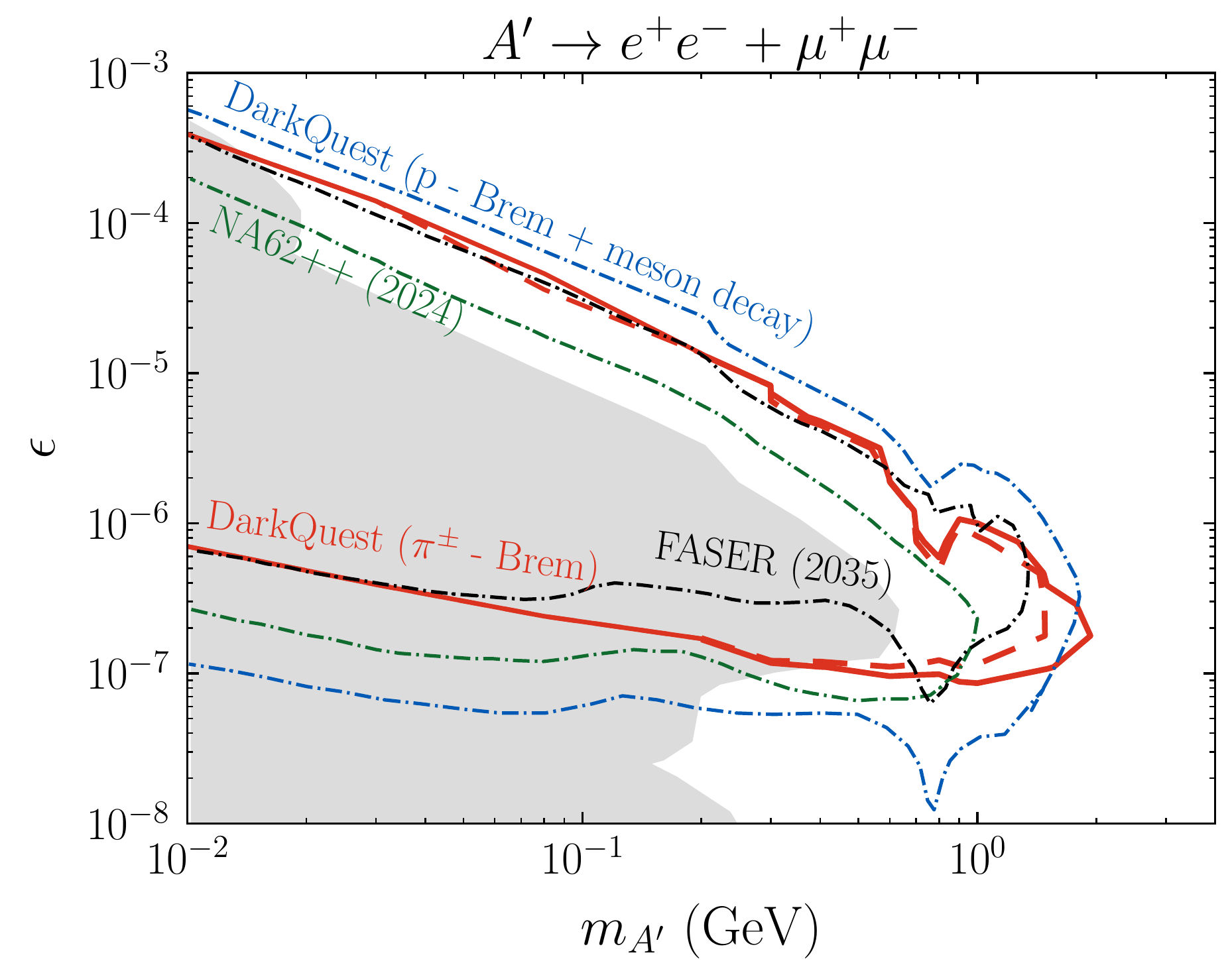}
    \caption{
    \textbf{Left:} The sensitivity of SpinQuest to $A'$ from charged pion bremsstrahlung. The red contours (solid for $M_T = M_N$, dashed for $M_T = M$) correspond to 10 muon signal events for the 5--6 m fiducial decay region in SpinQuest with $1.44 \times 10^{18}$ POT, with the width of the shaded region representing the modeling uncertainty for the nucleon distribution in the iron nucleus. The gray shaded region displays the parameter space that has been ruled out by previous beam dump experiments. \textbf{Right:} The sensitivity of SpinQuest to charged pion bremsstrahlung for $A' \rightarrow e^+ e^-, \ \mu^+ \mu^-$ decays. The red contours (solid for $M_T = M_N$, dashed for $M_T = M$) correspond to 10 lepton signal events for the 5--12 m fiducial decay region for the proposed DarkQuest upgrade with $10^{20}$ POT. We compare this curve to the projected sensitivity of DarkQuest to $A'$ from proton bremsstrahlung and neutral meson decays with corresponding lepton signal events (blue dot-dashed)~\cite{Berlin:2018pwi}. We have also included projected sensitivity contours from NA62++ (green dot-dashed) and FASER (black dot-dashed) \cite{Apyan:2022tsd}. 
    }
    \label{fig:sensitivity}
\end{figure*}

To simulate dark photon production, we sample $N_0$ pions from the binned energy spectrum in Fig.~\ref{fig:piondist}. For $i = 1, ..., N_0$, we denote $E_{\pi,i}^0$ as the initial energy of the charged pion. We assume all pions are produced purely in the forward direction and have negligible deflection due to multiple scattering. The energy of each pion as it steps a length $\Delta l$ through the beam dump is then 
\begin{equation}
    E_{\pi,i}(l_j) = E_{\pi,i}^0 \exp(-l_j / l_{\pi}),
\end{equation}
where $l_j = j \Delta l $ for $j = 1,...,(L / \Delta l)$ and $l_{\pi}$ is the pion interaction length, equal to $20.41$ cm in iron. For each dark photon produced from the pion with differential cross section $d\sigma/dx$, where $x = E_{A'} / E_{\pi,i}(l_j)$ is the $A'$ energy fraction, we bin $x$ into $m$ bins with left endpoints $x_k$, $k = 1, \dots m$. Then the number of dark photons produced is
\begin{equation}
    \begin{split}
    N_{A'}(m_{A'}, \epsilon) = \frac{N_\pi}{N_0} \sum_{i = 1}^{N_0} \sum_{j = 1}^{L / \Delta l} \sum_{k = 1}^{m}  \Delta l \\
    \times \int_{x_k}^{x_{k+1}} dx \frac{d \sigma}{dx}(m_{A'}, \epsilon, x, E_{\pi,i}(l_j))
    \end{split}
\end{equation}
The result is a distribution of dark photons labeled by energies $E_{A'}$ and the location, $l_{A'}$ it was produced in the FMAG. Recall that we have already integrated over the emission angle $\theta$ of the dark photon, conservatively assuming that all dark photons are produced at $l_j = 0$ where the geometric acceptance of the experiment is the smallest.

Finally, we use (\ref{eq:Nsignal}) to calculate the number of dark photons that decay within the fiducial decay region. We consider the total decay width into muons for the current SpinQuest setup and electrons and muons for the DarkQuest upgrade. The dark photons must survive the length of the beam dump and decay in the fiducial decay region. We require that the leptons must reach station 3, so we apply a momentum cut of $p_{\rm min} = 10$ GeV to the dark photon to ensure that it is sufficiently boosted. We then isotropically decay the dark photons in the $A'$ rest frame within the fiducial decay region, boost back to the lab frame, and apply a kick of $\Delta p_T \simeq 0.4 \text{ GeV} \times (\Delta z / 3 \text{ m})$ from the KMAG. Only the decay products that are still within the angular resolution of station 3 are considered signal events.
 
\subsection{Estimated sensitivity to dark photons}

We now present our estimates of the sensitivity of SpinQuest/DarkQuest to the leptonic decay of dark photons produced from secondary charged pion bremsstrahlung. In Fig.~\ref{fig:sensitivity} (left), we show the 10-event sensitivity to $A' \rightarrow \mu^+ \mu^-$ for the 5--6 m decay region with the expected SpinQuest luminosity. The parameter space already ruled out by previous beam dump experiments~\cite{Riordan:1987aw,Bross:1989mp,Bjorken:1988as,Batell:2014mga,Marsicano:2018krp,Blumlein:2011mv,Blumlein:2013cua,Gninenko:2012eq} and supernova 1987A~\cite{Chang:2016ntp} is shown in shaded grey. Remarkably, even accounting for nuclear modeling uncertainty, dark photon production \emph{exclusively} from  charged pion bremsstrahlung allows SpinQuest to probe new parameter space even prior to the DarkQuest upgrade.

To explore the prospects of DarkQuest, we show the sensitivity to $A' \rightarrow e^+ e^-, \ \mu^+ \mu^-$ in purple in Fig.~\ref{fig:sensitivity} (right). To directly compare to previous work, we also show the projected sensitivity curves from Ref.~\cite{Berlin:2018pwi} for DarkQuest from the meson decay and primary proton production channels (blue dot-dashed), which correspond to 10 signal events for the upgraded proton beam luminosity.  The secondary charged pion channel alone yields comparable sensitivity to the projected reach of FASER (black dot-dashed), and depending on a full treatment of the nucleuon distributions in the nucleus, could extend the reach of DarkQuest out to $m_{A'} \sim 2 \ {\rm GeV}$. Indeed, the enhancement in the number of dark photons produced with larger POT is more pronounced at large masses, and the reach can extend to larger masses because the signal is no longer event-rate limited. The projected mass reach of the charged pion bremsstrahlung channel also exceeds the NA62++ projections (green dot-dashed); to our knowledge, no other experiment (other than the proposed SHiP~\cite{SHiP:2015vad}) would have sensitivity to dark photons in this mass and coupling range. While there is an extension of mass reach, naively Fig.~\ref{fig:sensitivity} (right) might imply that the impact of dark photon bremsstrahlung by secondary pions has only a very minor impact, especially if the correct kinematics for nuclear scattering makes the effective target mass closer to that of a nucleon than the whole nucleus. However, our calculation must be viewed as the motivating first step in obtaining the full dark photon signal contribution of secondary pions; we discuss next steps in the Conclusion below.

\section{Conclusion}
\label{sec:conclusions}

In this paper, we have shown that a previously-neglected production channel for dark photons, bremsstrahlung from secondary charged pions produced from a primary proton beam, can extend the sensitivity of proton beam dump experiments to kinetically-mixed dark photons. 
Our analysis shows that the expected SpinQuest dataset should have sensitivity to dark photons beyond existing constraints, even allowing for the fact that only the muon decay mode is visible. Over the next decade, the DarkQuest upgrade will allow more decay channels and higher luminosity, pushing sensitivity into the multi-GeV regime.

We have made a number of simplifiying assumptions that underestimate the true reach; most importantly, we have neglected the $\rho$ resonance in pion-nucleon scattering as well as dark photons produced from quark-level scattering at momentum transfers exceeding the cutoff of the chiral Lagrangian. As Figure.~\ref{fig:NAproduced_ep1e-6} demonstrates for proton bremsstrahlung vs. Drell-Yan type processes, quark-level processes become dominant for dark photon masses in the GeV-range and beyond. Therefore, the fact that the dark photon production rate from pion bremsstrahlung dominates over proton bremsstrahlung above $\sim$ GeV regardless of the assumed nuclear kinematics implies that pion-initiated quark-level processes will in fact signficantly increase the dark photon mass reach compared to our current estimates. We will study these processes in an upcoming analysis in order to find out just how far in large $m_{A'}$ the reach can go. 

We can already get some intuition for the expected rate increase from quark-level scattering by comparing the estimated proton-initiated Drell-Yan rate to the full rate including proton PDFs in Figure~\ref{fig:NAproduced_ep1e-6}. In Ref.~\cite{Berlin:2018pwi}, the Drell-Yan cross section was parametrically estimated as $\sigma_{{\rm DY}} \sim \alpha \epsilon^2/m_{A'}^2$, which would lead to $\mathcal{O}(1)$ dark photon with $m_{A'} = 1 \ {\rm GeV}$ and $\epsilon = 10^{-6}$ for $10^{18}$ POT. However, incorporating the full proton PDFs yielded a rate orders of magnitude larger, as can be seen from Fig.~\ref{fig:NAproduced_ep1e-6}. For pion-initiated Drell-Yan, one obvious advantage is that the (valence) antiquark parton distribution functions in the pions are $\mathcal{O}(1)$ at $x_{\bar{q}} \sim 0.5$ at the momentum transfers in question~\cite{Novikov:2020snp}, unlike the case of the proton where antiquark PDFs are from sea quarks and peak at $x_{\bar{q}} \ll 1$. If pion-initiated 
Drell-Yan processes scale similarly to the proton-initiated process, then it is reasonable to expect great enhancements in the resulting dark photon mass reach at DarkQuest. 

Our work has profound consequences for the detector and analysis upgrade path from SpinQuest to DarkQuest. Without including dark photon production from secondary pions, dark photon masses significantly beyond a GeV are essentially inaccessible, owing to the sharp drop off in proton bremsstrahlung production past the proton mass. 
We have argued that dark photon production from secondary pion interactions exceeds other channels for $m_{A'} \gtrsim$~GeV,
opening up the multi-GeV dark photon regime.
Dark photons with such masses decay dominantly to hadrons, meaning that sensitivity could be further enhanced by including hadronic $A'$ decays in the dark photon search at DarkQuest. 
We therefore call for an urgent investigation into possible optimizations of the proposed DarkQuest upgrade of SeaQuest in light of this new information, to ensure that hadronic long-lived particle decays could be discovered with sufficiently low background and obvious possibilities for the discovery of new physics are not missed.

\section*{Acknowledgments}

We thank Patrick Draper, Philip Harris, Christian Herwig, Jaki Noronha-Hostler, Jen-Chieh Peng, Matt Sievert, and Nhan Tran for discussions, and Josh Foster for collaboration in the early stages of this work. RN and YK also thank Asher Berlin for assistance with Pythia, Diana Forbes for cross-checks with CalcHEP, and Yi-Ming Zhong for invaluable assistance with MadGraph. RN is supported by the National Science Foundation Graduate Research Fellowship Program under Grant No. DGE – 1746047. YK is supported in part by DOE grant DE-SC0015655. 
The research of DC was supported in part by a Discovery Grant from the Natural Sciences and Engineering Research Council of Canada, the Canada Research Chair program, the Alfred P. Sloan Foundation, the Ontario Early Researcher Award, and the University of Toronto McLean Award.

\appendix
\section{$2 \to 3$ phase space and kinematic limits}
\label{app:CS}

Here we present the derivation of the required 3-body phase space integral in a form convenient for obtaining differential cross sections in the lab frame, following the procedure in \cite{Liu:2016mqv}. This is an elementary calculation, but we show it here for pedagogical purposes and to correct some typos in the literature. 

The Lorentz-invariant 3-body phase space measure for $\pi (p) N (p_i) \to \pi (p') N (p_f) A' (k)$ is
\begin{equation}
\begin{split}
    d \Pi_3 =\frac{1}{2^8 \pi^5} \frac{d^3 \ppr d^3 \pf d^3 \kk}{E' E_f E_k} \delta^4(p+p_i-p_f-p'-k),
\end{split}
\end{equation}
where we define 4-momenta as follows: $p = (E_\pi, \pp_\pi)$, $p' = (E', \ppr)$, $p_i = (E_i, \mathbf{p}_i)$, $p_f = (E_f, \pf)$, and $k = (E_k, \kk)$. As discussed in the main text, we take the mass of $N$ to be $M_T$, which can vary between $M$ and $M_N$ depending on the (unmodeled) distribution of nucleons in the nucleus.

We can collapse the delta function over the 3-momenta by doing the integral over $\ppr$. This sets 
\begin{equation}
    \ppr = \pp_\pi + \textbf{p}_i - \pf - \kk.
\end{equation}
We work in the lab frame where $\textbf{p}_i = 0$. Following the conventions of Ref.~\cite{Liu:2016mqv}, we also define $\q = -\pf$ as the (negative of the) momentum transfer to the nucleus, and let
\begin{equation}
 \V = \kk - \pp_\pi = \q - \ppr.
\end{equation}
Then we can set the energy of the outgoing pion to
\begin{equation}
\begin{split}
    (E')^2 &= m_{\pi}^2 +| \q - \V|^2  \\
    &= m_{\pi}^2 + |\q|^2 + |\V|^2 - 2 |\q| |\V| \cos \theta_q,
\end{split} 
\end{equation}
where $\theta_q$ is the angle between $\q$ and $\V$.
Changing variables from $\pf$ to $\q$, and letting $\phi_q$ be the azimuthal angle of $\q$ with respect to $\V$, the phase space measure is now
\begin{equation}
\begin{split}
     d \Pi_3 = \frac{1}{2^8 \pi^5} \frac{d^3 \kk}{E' E_f E_k} |\q|^2 d|\q| \, d\cos \theta_q \, d \phi_q  \, \\
     \times \, \delta(E_\pi +M_T-E_f-E'-E_k).
\end{split}
\end{equation}
To integrate the remaining energy delta function\footnote{At this step there is a typo in \cite{Liu:2016mqv}, which states that the delta function is used to integrate out $Q$, but is actually used to integrate out $\cos\theta_q$.}, we treat the argument of the delta function as a function of $\theta_q$:
\begin{equation}
\begin{split}
    f(\cos \theta_q) & = E_\pi +M_T-E_f -E_k \\ 
    & - \sqrt{m_{\pi}^2 + |\q|^2  + |\V|^2 - 2 |\q| |\V| \cos \theta_q},
\end{split}
\end{equation}
with Jacobian $f'(\cos \theta_q) = |\q||\V|/E'$. The $E'$ cancels the $E'$ in the phase space measure, and upon integrating over $\cos \theta_q$ we obtain
\begin{align}
    d \Pi_3 = \frac{1}{2^8 \pi^5} \frac{|\q| d|\q| d \phi_q}{E_f E_k |\V|} \, |\kk|^2 d|\kk| \, d\cos \theta \, d \phi,
\end{align}
where we have switched to spherical coordinates for the $A'$ momentum, with $\theta$ and $\phi$ the polar and azimuthal angles of $\kk$ with respect to the incoming pion.

We now want to change variables from $|\kk|$ to energy fraction $x \equiv E_k/E_\pi$. To do this we use
\begin{equation}
    E_k = \sqrt{|\kk|^2 + m_{A'}^2} = x E_\pi \quad \xRightarrow \quad d|\kk| = \frac{E_k E_\pi}{|\kk|} dx.
\end{equation}
Plugging this into the phase space integral and exploiting the azimuthal symmetry of the problem to perform the integral over $\phi$, we have
\begin{align}
    d \Pi_3 = \frac{1}{2^7 \pi^4} \frac{E_\pi |\kk| |\q|}{E_f |\V|} \, dq \, d\phi_q \, dx \, d\cos \theta.
\end{align}
Finally, we change variables from $|\q|$\footnote{Another typo in \cite{Liu:2016mqv}, where the change of variables should be from $|\q|$ to $t$ rather than $\cos \theta_q$ to $t$.} to
\begin{equation}
    t = -(p_i - p_f)^2 = -2 M_T^2 + 2 M_T \sqrt{M_T^2 + |\q|^2},
\end{equation}  
where the minus sign ensures that $t$ is positive-definite.
The Jacobian is
\begin{equation}
    d|\q| = \frac{E_f}{2 M_T |\q|} dt.
\end{equation}
Then the final phase space measure involves 4 integration variables: 
\begin{equation}
    d \Pi_3 = \frac{1}{2^8 \pi^4} \frac{E_\pi |\kk|}{M_T |\V|} dt\, d\phi_q\, dx\, d \cos \theta,
\end{equation}
where
\begin{align}
    |\kk| & = \sqrt{(x E_\pi)^2 - m_{A'}^2}, \\
    |\V| & = \sqrt{(1 - \cos \theta^2) |\kk|^2 + (|\kk| \cos \theta - \sqrt{E_\pi^2 - m_\pi^2})^2}.
\end{align}
Note that no further analytic progress is possible because the matrix element is in general a nontrivial function of $t$, $\phi_q$, $x$, and $\cos \theta$. Including the matrix element and the appropriate flux factor for the cross section in the lab frame,
\begin{equation}
    d\sigma = \frac{1}{4 E_\pi M_T} \mathcal{M}^{(p,n)}_{2 \rightarrow 3} d\Pi_3,
\end{equation}
yields Eq.~(\ref{eq:totalcs}) in the main text after summing incoherently over protons and neutrons in the nucleus.

\begin{figure}
    \centering
    \includegraphics[width = 0.47\textwidth]{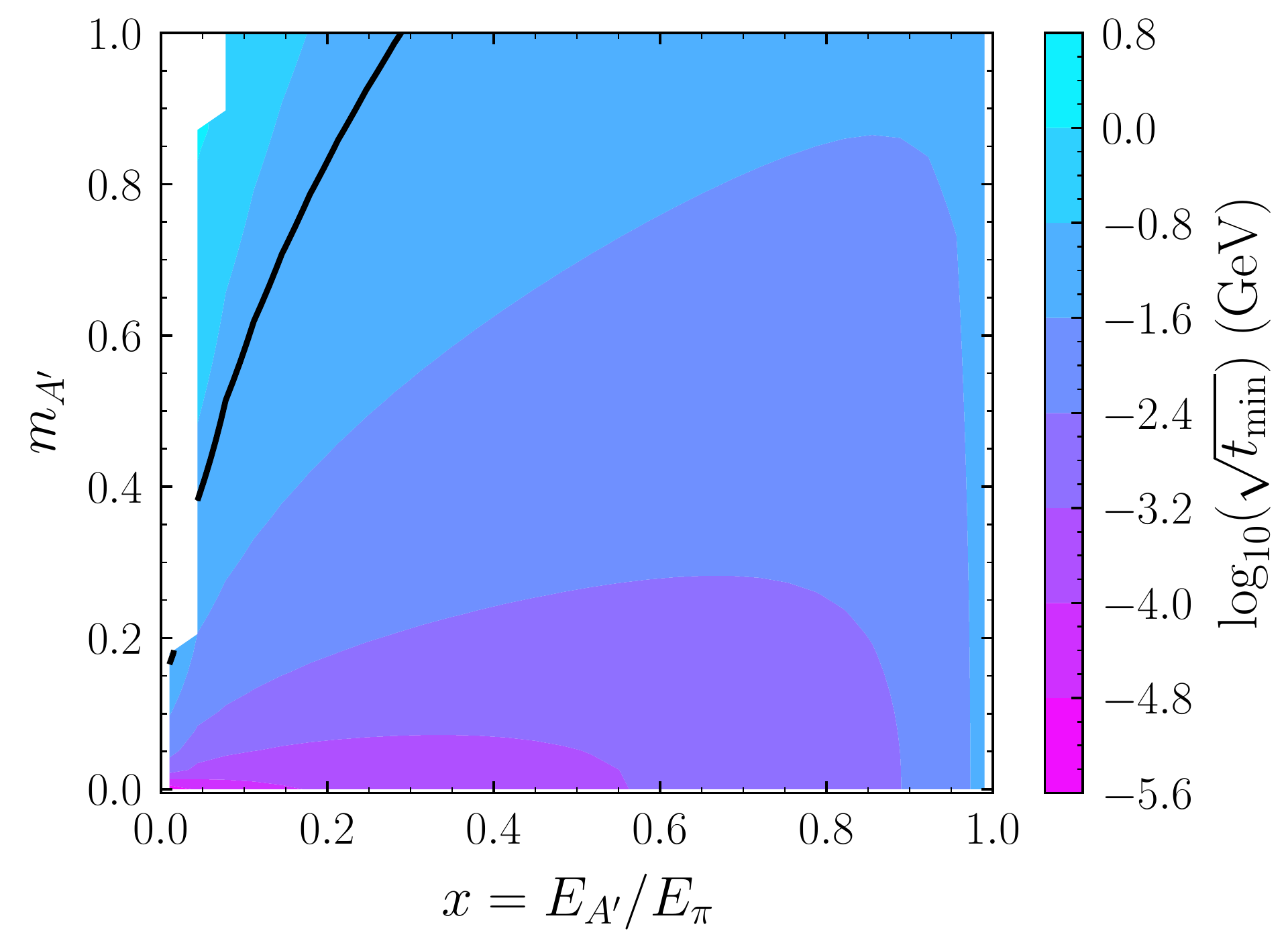}
    \caption{ The minimum momentum transfer, $\sqrt{t_{\rm min}}$, as a function of dark photon mass and energy fraction for a beam energy of $E_{\pi} = 20$ GeV and $M_T = M$. The momentum transfer is minimized at $m_{A'} = 0$ and $x = 0$, which is the SM photon limit. Since the cross section is proportional to the inverse of $t_{\rm min}$, the cross section is maximized in the SM photon limit, suppressing the QED process of dark photon bremmstrahlung. We also show the contour corresponding to the scale, $\sqrt{t} = f_{\pi}$, in black. }
    \label{fig:tmin}
\end{figure} 

We now consider the bounds on the phase space integration variables. First, the minimum energy of the $A'$ is when it is at rest in the lab frame, $E_{A'} = m_{A'}$, so the lower limit of $x$ is
\begin{equation}
    x_{\rm min} = \frac{m_{A'}}{E_\pi}.
\end{equation}
We can likewise find the upper bound of $x$ by considering the case where the $A'$ is collinear with the incoming pion and takes all of its kinetic energy, with the target nucleus remaining at rest. The initial 4-momentum of the system is $P_i = (E_{\pi}+M_T,0,0,\sqrt{E_{\pi}^2 - m_{A'}^2})$, and the final 4-momentum is $P_f = (m_{\pi}+M_T+E_{A'},0,0,\sqrt{E_{A'}^2 - m_{A'}^2})$. Conservation of 4-momentum yields 
\begin{equation}
    x_{\rm max} = \frac{2 E_{\pi} M_T - m_{A'}^2 - 2 M_T m_{\pi}}{2 E_{\pi}(M_T + m_\pi)}.
\end{equation} 

\begin{figure}[t]
    \centering
    \includegraphics[width = 0.47\textwidth]{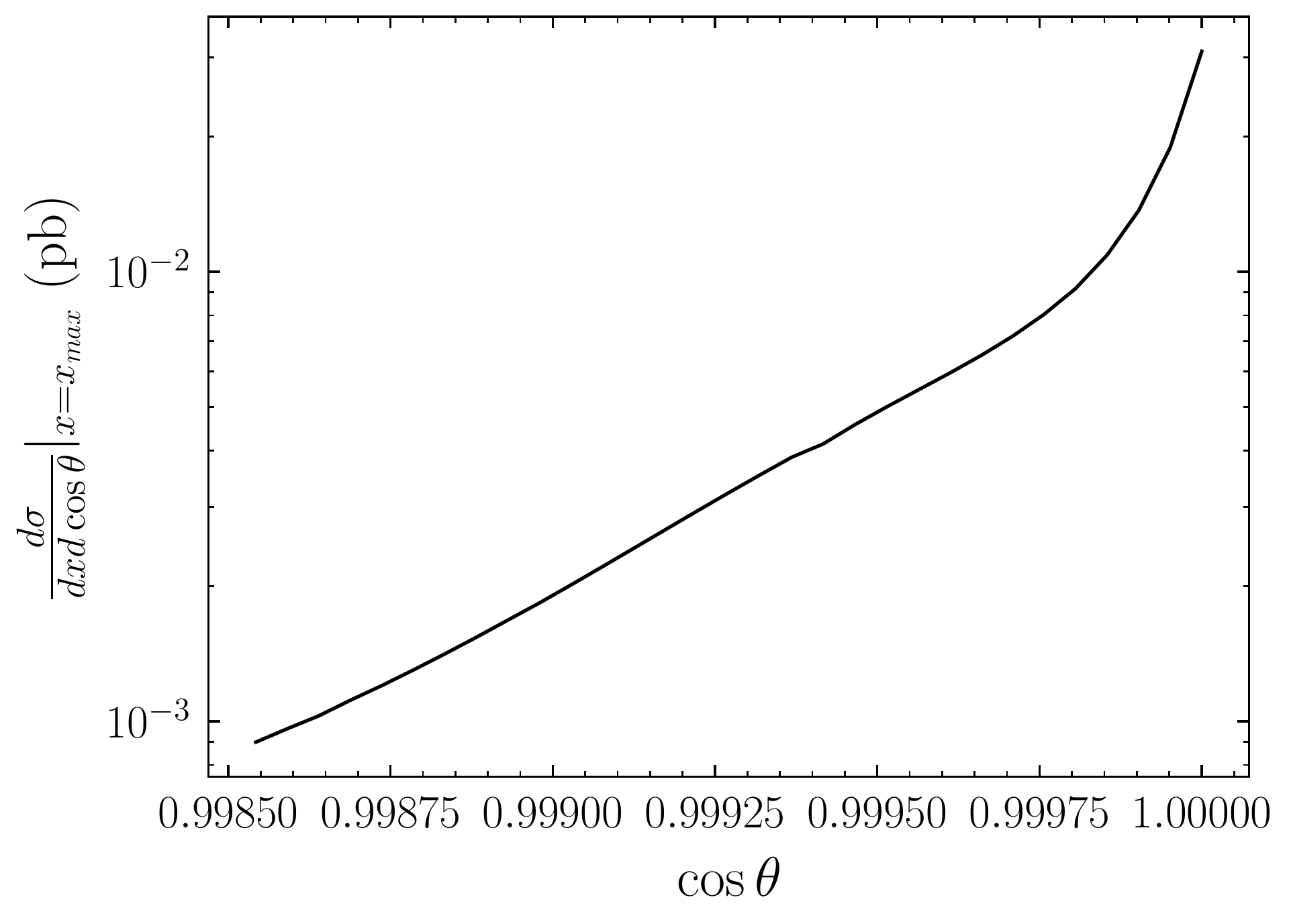}
    \caption{The differential cross section as a function of $\cos \theta$ at fixed $x = x_{\rm max}$ for a beam energy of $E_{\pi} = 30$ GeV and a dark photon with $m_{A'} = 1$ GeV and $\epsilon = 10^{-7}$.}
    \label{fig:thetadistribution}
\end{figure}

To determine the remaining integration bounds, we first find the bounds on $|\q|$ using energy conservation. For fixed $|\q|$ and $|\V|$, the minimum outgoing pion energy $E'_-$ is obtained when $\cos \theta_q = 1$ and must satisfy
\begin{equation}
    E'_-  \leq E_\pi + M_T - E_f - E_k.
\end{equation}
This implies 
\begin{equation}
\label{eq:qlim}
    m_{\pi}^2 + |\q|^2 + |\V|^2 - 2 |\q| |\V| \leq (E_{\pi} + M_T - E_f - E_k)^2,
\end{equation}
where $E_f = \sqrt{M_T^2 + |\q|^2}$. After simplifying, Eq.~(\ref{eq:qlim}) reduces to a quadratic equation for $|\q|$, with two roots $|\q|_\pm$. Then the bounds on $t$ are
\begin{equation}
\label{eq:tmin}
    t_\pm = 2M_T \left(\sqrt{M_T^2 +|\q|_\pm^2} - M_T\right).
\end{equation}
The lower limit of the $t$ integral is thus $t_{\rm min} = t_-$. Fig.~\ref{fig:tmin} shows a contour plot of $\sqrt{t_{\rm min}}$ as a function of $x$ and $m_{A'}$ for $E_\pi = 20$ GeV. As discussed in the main text, to ensure the validity of chiral perturbation theory, we set the upper limit at 
\begin{equation}
    t_{\rm max} = {\rm min} \{ (4\pi f_\pi)^2, t_+ \}.
\end{equation}
Finally, when the discriminant of the $|\q|$ equation vanishes, we have $t_+ = t_-$ and the phase space integral also vanishes since the bounds of the integral are degenerate. This implies a maximum value for $\theta$ as a function of $x$,
\begin{equation}
    \cos \theta_+ =  \frac{2 x E_\pi(E_\pi + M_T) - 2M_T(E_\pi - m_\pi) - m_{A'}^2}{2\sqrt{(E_\pi^2 - m_\pi^2)((x E_\pi)^2 - m_{A'}^2)}},
\end{equation}
where larger $x$ implies smaller $\theta_+$ and hence more collinear emission. In practice, the angular emission is cutoff by the chiral cutoff. At small values of $\theta_+$, $t_{\rm min}$ becomes larger than $(4\pi f_\pi)^2$, prohibiting any dark photons to be emitted at large angles under chiral perturbation theory. In the parameter space relevant to SpinQuest, we take $\theta_{\rm max} = \text{min} \{\theta_{\chi}, 0.05 \text{ rad}\}$ such that the angular integral is determined by the either the experimental geometry or the allowed phase space. However, the cross section is insensitive to the large-angle limit of integration because it is sharply peaked at $\theta = 0$, as shown in Fig. \ref{fig:thetadistribution}. We have verified that with the unrestricted limits $t_\pm$ and $\theta_+$, the 3-body phase space volume reproduces the standard result in the center-of-mass frame~\cite{ParticleDataGroup:2016lqr}.

\section{Suppression of the QED scattering process}
\label{app:QEDsuppress}

As secondary charged pions traverse the FMAG, they scatter with atomic nuclei through both the electromagnetic and the strong force. In addition to the chiral perturbation theory calculation, one can consider dark photons produced through electromangetic scattering via exchange of a $t$-channel photon. However, the QED process is highly suppressed for the relevant kinematics, such that it can be neglected compared to the strong-force scattering.

The suppression comes about because there is a kinematic mismatch between the forward-scattering singularity of QED and the large momentum transfers required to produce massive dark photons. The QED cross section for $\pi^\pm N \to \pi^\pm N A'$ is parametrically
\begin{equation}
    \sigma_{\rm QED} \sim \frac{\epsilon^2 \alpha^3}{t_{\rm min}},
\end{equation}
where $t_{\rm min}$ is the smallest kinematically-allowed squared momentum transfer, defined in Eq.~(\ref{eq:tmin}). As shown in Fig.~\ref{fig:tmin}, for GeV-scale boosted $A'$s with $x \gtrsim 0.5$ (which are likely to result in detectable signals), $t_{\rm min}$ is on the order of $(50 \ {\rm MeV})^2$. On the other hand, as shown in Eq.~(\ref{eq:csparametrics}), the chiral perturbation theory cross section is parametrically
\begin{equation}
    \sigma_{\chi \rm{PT}} \sim \frac{\alpha \epsilon^2}{f_\pi^2},
\end{equation}
where there is no forward singularity at $t = 0$ but rather contact interactions at the scale $f_\pi$. Since $t_{\rm min}$ is on the same order as $f_\pi$, the ratio of the cross sections is parametrically
\begin{equation}
    \frac{\sigma_{\rm QED}}{\sigma_{\chi \rm{PT}}} \sim \alpha^2 \approx 10^{-4},
\end{equation}
which justifies neglecting the QED contribution. The QED cross section can be significant for $m_{A'} \lesssim m_\pi$ when $t_{\rm min}$ is smaller, but it is still suppressed compared to the dominant $\pi^0 \to \gamma A'$ production mode by 3-body phase space and an additional factor of $\alpha$.

\bibliography{references}

\end{document}